\documentclass[12pt,preprint]{aastex}

\shorttitle{Magnetorotational Instability around a Rotating Black Hole}
\shortauthors{Yokosawa et al.}
\begin{document}

\title{Magnetorotational Instability around a Rotating Black Hole}

\author{M. Yokosawa and T. Inui} 
\affil{Department of Physics, Ibaraki University, Mito, Japan 310-8512}

\begin{abstract}
The magnetorotational instability(MRI) in the Kerr spacetime is
 studied on a 3+1 viewpoint. The Maxwell's equations are expressed
 in a circularly orbiting observer's frame that co-rotates with matter
 in Keplerian orbits. The hydromagnetic equations are represented in a
 locally nonrotating frame(LNRF). There exist large proper growth rates
 in MRI around a rapidly rotating black hole. The large "centrifugal
 force" and the rapid variations of magnetic fields are caused by the
 rotation of spacetime geometry. As the result, in the extreme Kerr case
 the maximum proper growth rate at $r=r_{ms}$ becomes about twelve times
 as large as that in Schwartzshield case, where $r_{ms}$ is the radius
 of a marginally stable orbit. The unstable range of wavenumber expands
 according to the rotational speed of spacetime geometry. The over
 stable mode of the instability becomes remarkable when the circular
 motion of the disk is quasi-relativistic and the strength of magnetic
 field is so large that the Alfven velocity is $v_A \ge 0.1c$. When the
 waves of perturbations propagate in the radial direction, these waves
 oscillate and their amplitudes grow exponentially. This 
 instability is caused by the differential rotation of the circularly 
 orbiting frame. The universality of the local Oort $A$-value of the disk
 is discussed in curved spacetime. In the
 extreme Kerr geometry, the amplitude of the maximum growth rate in the dynamical shear instability (DSI) reaches to infinity at $r=r_{ms}$. The behaviors of maximum growth rates between MRI and DSI are remarkably different each other.
\end{abstract}


\section{Introduction}

The various intensive phenomena, x-ray or $\gamma$-ray emission, jet,
etc. are found in many active objects. Some intense $\gamma$-ray bursts
(GRBs) are believed to arise from the hot, dense matter accretion onto a
rotating black hole \citep{PWF,MPN,KM,YUA04}. The viscosity is the important
physical quantity to balance the heat in the disk and to determine the
dynamics of the accretion. The magnetohydrodynamical turbulent viscosity
is driven by the magnetorotational instability (MRI)
\citep{BH91,HB91,HGB95,B95,MT95,S96,BH98}. We study here the
MRI on the basis of the general relativistic magnetohydrodynamics.

The MRI in the Kerr metric had been considered by \citet{G04}. He had
analyzed the dynamical shear instability(DSI) for circularly orbiting
particles. The maximum growth rate in MRI was then discussed
according to the analogy between DSI and MRI. It had been previously
shown by \citet{BH92} that in a Newtonian case the both maximum growth
rates in MRI and DSI coincide each other. However its 
consistency is not always satisfied in general spacetime. The MRI in
curved spacetime should be investigated on the general
relativistic Maxwell's equations and hydrodynamics, and then be examined
for the consistency of the maximum growth rate between the two types of
instabilities. 

The generation of MRI is caused by the two processes; the strengthen of
magnetic field due to the differential rotations of the disk, and the
transfer of angular momentum by magnetic stress \citep{BH98}. The
rotation of the spacetime geometry exerts the characteristic effects on the
kinematics and on the induction of magnetic field.
In the rotating spacetime geometry, there appears "the gravitomagnetic
force" which is an inertial force in 3+1 point of view \citep{TPM86,
P01}. The gravitomagnetic force depends on the gradient of the
geometrical angular velocity, ${\nabla}\omega$. The gravitomagnetic
force exerts a torque on a spin outside a black hole \citep{TPM86} and
strengthens the centrifugal force. The increased radial velocity then induces
the large variation of magnetic fields which transfer the angular momentum. The rotating velocity of the disk becomes extremely large in the
vicinity of a rapidly rotating black hole. The changing rate of 
magnetic field due to the differential rotation of the disk is then largely
increased. The generation of MRI may be very active around a rapidly rotating
black hole.

It is useful to introduce the 3+1 formulation of general relativity
which is excellent to make clear the physics on the stability. The
evolutions of the perturbations are then simply evaluated by the
analogous terms to the Newtonian dynamics and to the Maxwell equations
in flat spacetime. We introduce the Keplerian orbiting observer's frame
(orbiting frame) to evaluate the electromagnetic fields perturbed by the 
motion of fluid. On the other hand the dynamic equations of fluid
interacting with magnetic fields are analyzed in the locally non-rotating
frame(LNRF)\citep{BPT} since this frame becomes an inertial frame at the
far distance 
from a black hole. Thus the dispersion relation expressed in these
frames approaches to a Newtonian one at the far distance.

When the Maxwell equations are described in the orbiting frame, there
appear explicitly the electromagnetic fields generated by the frame moving
. When the rotating velocity of the 
frame is relativistic, the induction of magnetic field by the
differential rotation of the orbiting frame becomes efficient. This
generation of 
magnetic field causes an overstable mode of instability. We investigate
here the characteristic behaviors of this new mode in MRI.

The intensive phenomena in active objects may be produced by the
nonstationary magnetohydrodynamical processes in the accretion 
disk around a rotating black hole. We are interesting in the spatial
distribution of a growth rate of MRI. \citet{YUA04} had
represented the angular velocity of circular orbits for a geometrically thick
disk, $\Omega(r,z)$. By using this angular velocity $\Omega(r,z)$, we
investigate the distribution of 
the maximum growth rate $\Gamma_{max}(r,z)$over the space around a
rotating black hole. 

When the geometry of spacetime rotates, it is important to select the
suitable frame for the analysis of the instability. In the
Boyer-Lindquist coordinate frame (BLCF) the dragging of inertial frames
becomes so severe that the $t$ coordinate basis vector
$(\partial/\partial t)$ goes spacelike at the static limit $r_0$(outer
boundary of the ergosphere). In LNRF the frame dragging effects of the hole's rotation are canceled out. \citet{G04} investigated the DSI in
BLCF. We reanalyze the upper bound to 
the growth rate of DSI in LNRF by using the general relativistic Hill
equations. We then examine the asymptotic behaviors of the growth rates 
for $r \to r_{ms}$ when the frame is exchanged from BLCF to LNRF. We
inquires the suitable frame to evaluate the instability, MRI or DSI. The
universality of local Oort 
$A$-value in the shear instability is then examined by comparing the
dispersion relations for DSI and MRI in the suitable frame. 

In the next section we shall describe the Maxwell's equations in the
orbiting frame and the magnetohydrodynamic equations in LNRF. The
dispersion relation and the properties of growth rates in curved
spacetime are analyzed in the section 3. The universality of the 
growth rate of the shear instability in general, gravitational
field is discussed in section 4.

\section{Basic Equations}

The preferred frame describing the basic equations is selected
definitely to evaluate the perturbations. The LNRF becomes an inertial
frame at the infinite distance from a black hole. Thus the
characteristic frequencies in dynamics, e.g., epicyclic frequency, are
definitely expressed in LNRF. The perturbations of hydrodynamics are
represented in LNRF. The
perturbations of magnetic field exerting on the fluid are simply
evaluated in the frame co-moving with the fluid in a Keplerian orbit
\citep{P01}. The both frames, LNRF and orbiting frame, are combined with the Lorenz transformation.

We choose units with $G = c = 1$. The metric of spacetime, in terms of
Boyer-Lindquist 
time $t$ and any arbitrary spatial coordinates $x^j$, has the form
\begin{equation}
ds^2 = -\alpha^2dt^2 + g_{jk}(dx^j+\beta^jdt)(dx^k+\beta^kdt).
\end{equation}
The metric coefficients $\alpha, \beta^j, g_{jk}$ are the lapse, shift,
and 3-metric 
functions. Theses functions are given by
\begin{eqnarray}
\alpha &=& \Bigl(\frac{\Sigma\Delta}{A}\Bigr)^{1/2}, \qquad 
g_{rr}=\frac{\Sigma}{\Delta}, \qquad g_{\theta\theta}=\Sigma, \qquad 
g_{\varphi\varphi}=\frac{\sin^2{\theta}A}{\Sigma} \nonumber \\  
\beta^\varphi &=&  -\omega = -\frac{2Mar}{A}, \  \  \quad \qquad \qquad \beta^j=g_{jk}=0 \qquad {\rm 
for\ all\ other}\ j{\rm\ and}\ k, \nonumber
\end{eqnarray}
where $M$ is the mass of the black hole, $a$ is its angular momentum per unit mass, 
and the functions $\Delta,\Sigma,A$ are defined by
\begin{eqnarray}
\Delta = r^2 - 2Mr + a^2, \qquad \Sigma=r^2+a^2\cos^2{\theta}, \qquad 
A=(r^2+a^2)^2-a^2\Delta\sin^2{\theta}.  \nonumber
\end{eqnarray}

We introduce a set of local observers who rotate with the Keplerian 
angular velocity. Each observer carries an orthonormal
tetrad. For the 
Keplerian orbiting observer with a coordinate angular velocity of $\Omega$,
its world line is $r=$constant, $\theta=$constant, $\varphi=\Omega t$+
constant. The observer in the LNRF who rotates with the angular velocity $\omega$
measures the linear velocity of a particle moving in a Keplerian 
orbit as $v_{(\varphi)}= dx^{(\varphi)}/dx^{(0)} =\alpha^{-1}
\sqrt{g_{\varphi\varphi}}(\Omega  - \omega)$ where $x^{(0)}$ and
$x^{(\varphi)}$ are the time and spatial coordinates in LNRF. Its
Lorentz factor is ${\gamma}=(1- 
v_{(\varphi)}^2)^{-1/2}$. We adopt the orbiting frame with the set of
it's basis vectors: 
\begin{eqnarray}
{\bf e}_{\hat 0} &=& \gamma e^{-\nu}\big(\frac{\partial}{\partial 
t}+\Omega\frac{\partial}{\partial \varphi}\big), \qquad\qquad 
{\bf e}_{\hat\varphi} = \gamma \Bigl(e^{-\psi} 
\frac{\partial}{\partial 
\varphi}+ v_{(\varphi)}e^{-\nu}(\frac{\partial}{\partial 
t}+\omega\frac{\partial}{\partial \varphi})\Bigr), \\ 
{\bf e}_{\hat r} &=& e^{-\mu_1}\frac{\partial}{\partial r},\qquad\qquad \qquad\qquad \
{\bf e}_{\hat \theta} = e^{-\mu_2}\frac{\partial}{\partial 
\theta},
\end{eqnarray}
where $e^\nu, e^\psi, e^{\mu_1}, e^{\mu_2}$ are difined by 
\begin{equation}
e^\nu=\alpha, \qquad
e^{2\psi}=g_{\varphi\varphi},\qquad e^{2\mu_1}=g_{rr}, \qquad e^{2\mu_2}=g_{\theta\theta}.
\end{equation}

The corresponding orthogonal basis of one-forms is
\begin{eqnarray}
{\vec \omega}^{\hat{t}} &=& \hat{\alpha}{\rm\bf d}t 
- v_{(\varphi)}\gamma e^\psi {\rm\bf d}\varphi, \qquad\qquad
{\vec \omega}^{\hat{\varphi}} = -\Omega \gamma e^\psi {\rm\bf d}t 
+  \gamma e^\psi
{\rm\bf d}\varphi \nonumber \\
{\vec \omega}^{\hat{r}} &=& e^{\mu_1}{\rm\bf d}r,\quad \qquad\qquad\qquad \qquad 
{\vec \omega}^{\hat{\theta}} = e^{\mu_2}{\rm\bf d}\theta. 
\label{eq: one-form}
\end{eqnarray}
Here $\hat{\alpha}$ is defined by $\hat{\alpha}=\alpha\gamma(1+ v_\omega v_{(\varphi)})$, where $v_\omega$ is the "rotating velocity of the geometry of spacetime", $v_\omega= \omega e^{\psi-\nu}$. 
The rotation one-forms, ${\omega}^{\hat{a}}_{\hat{b}}$, which allow one
to read off the connection coefficients $\Gamma^{\hat{a}}_{\hat{b}
\hat{i}}$ by ${\omega}^{\hat{a}}_{\hat{b}} = \Gamma^{\hat{a}}_{\hat{b}
\hat{i}}{\vec \omega}^{\hat{i}}$. Hereafter the components of the tensor
in the orbiting frame are depicted by the superscript with hat. The
connection coefficients $\Gamma^{\hat{a}}_{\hat{b}
\hat{i}}$ are represented in Appendix A. 

The Maxwell equations for the electromagnetic field $F^{\hat\mu\hat\nu}$ produced by a four current density $J^{\hat\mu}$ are 
\begin{eqnarray}
F^{\hat\mu\hat\nu}_{;\hat\mu} = - J^{\hat\nu}, \qquad \qquad F_{\hat\mu\hat\nu;\hat\lambda} + F_{\hat\lambda,\hat\mu;\hat\nu} + F_{\hat\nu\hat\lambda;\hat\mu} = 0.
\end{eqnarray}
In the orbiting frame they are represented as follows:
\begin{eqnarray}
\frac{\partial {{\rm \bf B}}} {\partial{\hat\tau}} &+&
\frac{1}{\hat\alpha}  {{\nabla}}  \times ({\hat\alpha}
{{\rm \bf E}} ) \nonumber \\
&-& {{\rm \bf n}}_{\hat \varphi} e^{b}v_\Omega 
{{\rm \bf B}} \cdot {{\nabla}} \ln\Omega - e^{b}{\rm\bf  v}_{(\varphi)} \cdot {\rm\bf B} {{\nabla}} \ln{\hat\alpha} - {\gamma}^2 v_{(\varphi)}v_\Omega {{\rm \it \bf E}} \times {{\nabla}} \ln{(\Omega / \hat\alpha )}
 = 0, \label{MXW01} \\
\frac{\partial {\rm \it \bf E}} {\partial{\hat\tau}} &-&
\frac{1}{\hat\alpha} \nabla  \times ({\hat\alpha}
{\rm \bf B} ) \nonumber \\
&-& {\rm \bf n}_{\hat \varphi} e^b v_\Omega 
{\rm \bf E} \cdot {\nabla} \ln{\Omega} - e^b {\rm\bf v}_{(\varphi)} \cdot {\rm \bf E} {\nabla} \ln{\hat\alpha} - {\gamma}^2 v_{(\varphi)}v_\Omega {\rm \bf B} \times \nabla \ln{(\Omega / \hat\alpha )}
 = - \frac{4 \pi}{c} {\rm \bf J}, \label{MXW02} \\
\nabla \cdot {\rm\bf E} &-& v_{(\varphi)} {\rm\bf E} \cdot{\nabla} \Omega + e^{b} {\rm\bf v}_{(\varphi)} \cdot ({\rm\bf {B}} \times \nabla) \ln{\hat\alpha} = \frac{4\pi}{c} J^{\hat 0}, \label{MXW03} \\
\nabla\cdot {\rm\bf B} &-& v_{(\varphi)}\vec{\rm\bf B} \cdot {\nabla} \Omega - e^b {\rm\bf v}_{(\varphi)} \cdot ({\rm\bf {E}} \times \nabla) \ln{\hat\alpha} = 0. \label{MXW04}
\end{eqnarray}
Here $d\hat\tau$ is the variation of time in the orbiting frame and it 
combins with the coresponding time $dt$ in the static frame by  
$d\hat\tau=\gamma^{-1}\alpha dt$.
The symbol ${{\rm \bf n}}_{\hat i}$ is the unit vector in
the $\hat i $ direction in the space of the orbiting frame. The
vectors ${\rm \bf B} = B^{\hat i} {{\rm \bf n}}_{\hat i}$ and ${\rm\bf E}
= E^{\hat i} {{\rm \bf n}}_{\hat i} $ are the magnetic and
electric fields with components $E^{\hat r} = F^{\hat 0 \hat r}, B^{\hat\varphi}= F^{\hat r \hat
\theta}$, and so on. The symbol ${\nabla}$ is
the covariant derivative operator, and ${\rm\bf v}_{(\varphi)}, \ v_\Omega$ and
$e^b$ are defined by ${\rm\bf v}_{(\varphi)}=v_{(\varphi)} {\rm \bf
n}_{\hat\varphi},\ v_\Omega=\Omega e^{\psi-\nu}$ and
$e^b={\gamma}^2(1+v_{(\varphi)}v_\omega)$. When the
angular velocity of the frame is taken to be the geometrical angular
velocity, $\Omega \to \omega$, the above expressions of Maxwell equations (\ref{MXW01})---(\ref{MXW04}) result in the forms described in LNRF \citep{TM82,TPM86}.

The Lorentz force is represented by 
\begin{eqnarray}
\frac{1}{c} {\rm \bf J}\times{\rm \bf B} = -\nabla \frac{(\hat\alpha B)^2}{8\pi\hat\alpha} + \frac{1}{4\pi} {\rm \bf B} \cdot \nabla (\hat\alpha{\rm \bf B}) + {\gamma}^2 v_{(\varphi)}v_\Omega \Bigl( -  \frac{B^2}{4\pi}\nabla +  \frac{\rm \bf B}{4\pi}{\rm \bf B}\cdot \nabla \Bigr) \ln(\hat\alpha /\Omega).
\end{eqnarray}

The energy-momentum tensor in LNRF, $T^{(i) (j)}$, is expressed with the four velocity of fluid $u^{(i)}$, the pressure $P$, and the density $\rho$ as
\begin{equation}
T^{(i)(j)} = (\rho + P)u^{(i)}u^{(j)} + \eta^{(i)(j)}P.
\end{equation}
Hereafter the components of a tensor in LNRF are expressed by the superscript with the parenthesis.  The tensor $\eta^{(i)(j)}$ is the Minkowskian metric. 
The Euler equations with the Lorentz force, $F^{(i)(\nu)} J_{(\nu)}$, are represented by
\begin{eqnarray}
\frac{\partial (\rho+P) u^{(0)} u^{(i)}} {\partial \tilde\tau} +
\frac{1}{\alpha} \tilde\nabla \cdot \alpha (\rho + P){\rm \bf u} u^{(i)} +
\frac{1}{\alpha} \tilde\nabla{\alpha P} 
+ \Gamma^{(i)}_{(\mu)(\nu)} T^{(\mu)(\nu)} = F^{(i)(\nu)}J_{(\nu)},  
\end{eqnarray}
where $\tilde\tau$ and $\tilde\nabla$ are the time and the  
covariant derivative operator in LNRF. The variation of time, $d\tilde\tau$, is
combined with $ d\hat\tau$ by $d\tilde\tau = \gamma d\hat\tau$. The Lorentz force
is related likewise as $F^{(i)(\nu)} J_{(\nu)}= \gamma
F^{\hat{i}\hat\nu}J_{\hat\nu}$. The term $\Gamma^{(i)}_{(\mu)(\nu)}
T^{(\mu)(\nu)}$ is the "inertial force" in the $(i)$ direction.
The forces for $(i) = (r), (\theta)$ are expressed in explicit forms:
\begin{eqnarray}
\Gamma^{(i)}_{(\mu)(\nu)} T^{(\mu)(\nu)} &=& T^{(0)(0)} {\tilde\nabla}^{(i)} \ln\alpha + T^{(0)(\varphi)} v_{\omega} {\tilde\nabla}^{(i)}\ln\omega - T^{(\varphi)(\varphi)} {\tilde\nabla}^{(i)} \psi \nonumber \\
&+& T^{(i)(j)} {\tilde\nabla}^{(j)}{\mu_i} - T^{(j)(j)} {\tilde\nabla}^{(i)}{\mu_j} .
\end{eqnarray}
Here the $(r)$ component of $ T^{(i)(j)} {\tilde\nabla}^{(j)}{\mu_i} -
T^{(j)(j)} {\tilde\nabla}^{(i)}{\mu_j}$ is $
T^{(r)(\theta)} {\tilde\nabla}^{(\theta)}{\mu_1} - T^{(\theta)(\theta)}
{\tilde\nabla}^{(r)}{\mu_2}$, 
and the $(\theta)$ component is 
$T^{(\theta)(r)} {\tilde\nabla}^{(r)}{\mu_2} - T^{(r)(r)} {\tilde\nabla}^{(\theta)}{\mu_1}$.
The $(\varphi)$ component of the "inertial force" is
\begin{equation}
\Gamma^{(\varphi)}_{(\mu)(\nu)} T^{(\mu)(\nu)} = (\rho +p)u^{(\varphi)} {\rm \bf u}\cdot \tilde\nabla \psi.
\end{equation}
This term is correspond to the "Coliori's force". 

The equation of continuity is 
\begin{eqnarray}
\frac{\partial \rho u^{(0)}} {\partial \tilde\tau} + \frac{1}{\alpha} 
{\tilde\nabla} \cdot {\alpha} \rho {\rm \bf u} = 0,
\end{eqnarray}

\section{Groth Rates of Magnetorotational Instability}
\subsection{Dispersion Relation}

The magnetohydrodynamical variables are expanded by the
infinitesimal quantities, ${\rm\bf v} = {\rm\bf v}_0 + \delta 
{\rm\bf v}, {\rm\bf B} ={\rm\bf B}_0 + \delta{\rm\bf B}$,
$\rho=\rho_0+\delta \rho {\ \  \rm and \ \ } P=P_0+\delta P$. We assume the
axisymmetric perturbations and 
use the local WKB approximation. The variables are unified to be the
components in the orbiting frame by the transformations of the
hydrodynamic equations from LNRF to the
orbiting frame. The linearized equations are then described by
\begin{eqnarray}
\nabla \cdot \delta {\rm \bf v} &=& 0, \\
\frac{\partial \delta v^{\hat r}}{\partial\hat\tau} &-& 2{\tilde\Omega}_{\hat r}\delta v^{\hat \varphi} + \frac{1}{\gamma\rho} \frac{\partial \delta P}{\partial x^{\hat r}} - \frac{\delta \rho}{\gamma \rho^2}\frac{\partial P}{\partial x^{\hat r}} \nonumber \\
&+& \frac{1}{4\pi\rho} \Bigl( B^{\hat\theta} \frac{\partial \delta B^{\hat\theta}}{\partial x^{\hat r}} - B^{\hat \theta} \frac{\partial \delta B^{\hat r}}{\partial x^{\hat\theta}} + 2B^{\hat\theta}\delta B^{\hat\theta} \gamma^2 v_{(\varphi)} v_\Omega \nabla^{\hat r} \ln{(\hat\alpha /\Omega)} \Bigr) = 0, \label{eq vr} \\
\frac{\partial \delta v^{\hat \theta}}{\partial\hat\tau} &-& 2{\tilde\Omega}_{\hat\theta} \delta v^{\hat \varphi} + \frac{1}{\gamma\rho} \frac{\partial \delta\rho}{\partial x^{\hat \theta}}  - \frac{\delta \rho}{\gamma \rho^2}\frac{\partial P}{\partial x^{\hat \theta}} =0, \\
\frac{\partial \delta v^{\hat \varphi}}{\partial\hat\tau} &+& {\vec\kappa} \cdot \delta {\rm\bf v} - \frac{1}{4\pi\rho} B^{\hat\theta} \frac{\partial \delta B^{\hat\varphi}}{\partial x^{\hat \theta}} = 0, \label{eq vf}\\
\frac{\partial \delta B^{\hat r}}{\partial \hat\tau} &-& {\rm\bf B} \cdot \nabla \delta v^{\hat r} =0, \\
\frac{\partial \delta B^{\hat \theta}}{\partial \hat\tau} &-& {\rm\bf B} \cdot \nabla \delta v^{\hat \theta} - B^{\hat\theta} \delta v^{\hat r} \gamma^2 v_{(\varphi)} v_\Omega \nabla^{\hat r} \ln{(\hat\alpha /\Omega)} =0, \\
\frac{\partial \delta B^{\hat \varphi}}{\partial \hat\tau} &-& {\rm\bf B} \cdot \nabla \delta v^{\hat \varphi} - \gamma^2 v_\Omega \delta {\rm\bf B} \cdot \nabla \ln{\Omega} =0, \label{Bf} \\
\frac{5}{3\rho} \frac{\partial \delta\rho}{\partial \hat\tau} &+& \delta v^{\hat r} \frac{\partial \ln{P\rho^{-5/3}}}{\partial x^{\hat r}} + \delta v^{\hat\theta} \frac{\partial \ln{P\rho^{-5/3}}} {\partial x^{\hat \theta}} =0,
\end{eqnarray}
where the coefficients $\vec{\tilde\Omega} = ({\tilde\Omega}_{\hat r}, {\tilde\Omega}_{\hat \theta} )$ and $\vec\kappa=(\kappa_{\hat r},\kappa_{\hat \theta})$ are defined by
\begin{eqnarray}
\vec{\tilde\Omega} = \gamma (v_{(\varphi)} \nabla {\psi} - \frac{1}{2} v_\omega \nabla \ln \omega  ), \qquad {\rm and} \qquad
{\vec\kappa}= \gamma v_{(\varphi)} \nabla \ln{(\gamma^2 e^\psi v_{(\varphi)} )}. \label{Omeg}
\end{eqnarray}

With the spacetime dependence of perturbations,
$e^{i(k_{\hat r}x^{\hat r} + k_{\hat\theta} x^{\hat \theta} -\hat\Gamma \hat\tau)}$, the
explicit dispersion relation is derived as
\begin{eqnarray}
\frac{k^2}{k_{\hat\theta}^2} {\check\Gamma}^4 
&-& \Bigl[\kappa^2 + \Bigl(\frac{k_{\hat r}}{k_{\hat\theta}} N_{\hat\theta} - N_{\hat r} \Bigr)^2 \Bigr] {\check\Gamma}^2 
- ( \kappa^2 + \chi^2 ) ({\rm\bf k} \cdot {\rm\bf v}_A )^2 \nonumber \\
 &+& v_A^2 \gamma^2 v_{(\varphi)} v_\Omega \nabla^{\hat r} \ln{(\Omega/\hat\alpha)} \Bigl(2 \gamma^2 v_{(\varphi)} v_\Omega \nabla^{\hat r} \ln{(\Omega/\hat\alpha)} - i 3k_{\hat r} \Bigr) \frac{({\rm\bf k} \cdot {\rm\bf v}_A )^2} {(k_{\hat \theta} v_A)^2} {\check\Gamma}^2 = 0, \label{dis}
\end{eqnarray}
where
\begin{eqnarray}
{\check\Gamma}^2 = \hat\Gamma^2 - ({\rm\bf k} \cdot {\rm\bf v}_A )^2, \quad k^2=k_{\hat r}^2 + k_{\hat \theta}^2, \quad
N_{\hat i}^2 = - \frac{3}{5\gamma\rho} \frac{\partial P}{\partial x^{\hat{i}}}\frac{\partial \ln{P\rho^{-5/3}}}{\partial x^{\hat{i}}} \quad {\rm for \ \ }  \hat{i}= \hat{r}, \hat{\theta},
\end{eqnarray}
and
\begin{equation}
\kappa^2 = 2\Bigl({\tilde\Omega}_{\hat r} - \frac{k_{\hat r}}{k_{\hat\theta}} {\tilde\Omega}_{\hat\theta} \Bigr) \Bigl( \kappa_{\hat r} - \frac{k_{\hat r}}{k_{\hat\theta}} \kappa_{\hat\theta} \Bigr), \qquad 
\chi^2= - 2\gamma^2 v_\Omega \Bigl( {\tilde\Omega}_{\hat r} - \frac{k_{\hat r}}{k_{\hat\theta}} {\tilde\Omega}_{\hat\theta} \Bigr) \Bigl( \nabla^{\hat r} - \frac{k_{\hat r}}{k_{\hat \theta}} \nabla^{\hat\theta} \Bigr) \ln\Omega.
\end{equation}

\subsection{Growth Rates}

The dispersion relation (\ref{dis}) has the resemblant form to a
Newtonian case except the last term. 
When $v_A^2 \ll N^2=N_{\hat r}^2+N_{\hat\theta}^2$, i.e., the magnetic pressure is
negligibly small in comparison with the gas pressure, $\beta =
P_{gass}/P_{mag} \gg 1$, the last term containing $v_A^2 \gamma^2
v_{(\varphi)} v_\Omega$ can be ignored in 
(\ref{dis}). At the equatorial plane and far from
a black hole, the functions,  $\kappa^2$ and $\chi^2$, have the following forms:
\begin{equation}
\kappa^2 \to \frac{2\Omega}{r}\frac{(r^2\Omega)}{\partial r}, \qquad \chi^2 \to - \frac{\partial \Omega^2}{\partial \ln r}. 
\end{equation}
The equation (\ref{dis}) then coincides with a Newtonian dispersion relation 
\citep{BH91}. The equation (\ref{dis}) is the general dispersion relation of MRI for all radii and for all $a$.

In the Newtonian case, the maximum growth rate relative to a angular
velocity of a circular orbit, $\Gamma_{max}/\Omega$, has an universal
value in all radii and in all 
mass scales of a central object \citep{BH91}. Here we introduce the
angular velocity of a circular orbit measured in LNRF,
$\tilde\Omega=(\Omega-\omega)/\alpha$. At the next it is transferred
into the orbiting frame as $\hat\Omega= \gamma\tilde\Omega
=(\Omega-\omega)\gamma/\alpha$. It may be said that $-\hat\Omega$ is the
proper angular velocity of the geometry measured in the orbiting
frame. Let's consider first a simple case for $\beta \ll 1, \ k_{\hat
r}=B^{\hat r}=0$, and introduce the dimensionless variables with use of
a angular velocity  
$\hat\Omega$ as  
\begin{eqnarray}
{\bar\Gamma} = \hat\Gamma /\hat\Omega, \quad \bar{k} = ({\rm\bf k}\cdot{\rm\bf v}_A) /\hat\Omega, \quad
\bar{N}=N_{\hat r} /\hat\Omega , \quad \bar{\Omega}_{\hat r}= {\tilde\Omega}_{\hat r} /\hat\Omega, \quad
\bar\kappa = \kappa /\hat\Omega , \ \ {\rm and} \ \ \bar\chi = \chi / {\hat\Omega }^2.
\end{eqnarray}
Here it may be noticed that the dimensionless growth rate $\bar\Gamma$
coincides with that in LNRF,$\bar\Gamma =\hat\Gamma/\hat\Omega
=\tilde\Gamma/\tilde\Omega$, where $\tilde\Gamma$ is the growth rate
measured in LNRF, $\tilde\Gamma=-id\ln{\delta}/d\tilde\tau$. The disk is
assumed to be isothermal and the Brunt-V{\rm $\ddot a$}is{\rm $\ddot
a$}l{\rm $\ddot a$} frequency is taken to be ${\bar N}_{\hat\theta}^2 =
N_{\hat\theta}^2/ {\hat\Omega }^2=0.8$ and $\bar{N}_{\hat r}^2= N_{\hat
r}^2/ {\hat\Omega }^2 =0.01{\bar N}_{\hat \theta}^2$.  

The dispersion relation (\ref{dis}) will lead to instability for values of
$\bar k$ less than the critical value:
\begin{equation}
{\bar k}_{crit}^2= {\bar\chi}^2 - {\bar N}^2.
\end{equation}
The unstable mode of a perturbation has a maximum growth rate, ${\hat\Gamma}_{max}$, at the wavenumber, $\bar k = {\bar k}_{max}$ such as
\begin{equation}
{\hat \Gamma}_{max}^2 = - \frac{\chi^2}{4} \frac{(1-{\bar N}^2/{\bar\chi}^2)^2}{1+{\bar \kappa}^2/{\bar\chi}^2}, \label{Gmax} \qquad {\rm and} \qquad  
{\bar k}_{max}^2 =  \frac{\bar\chi^2}{4} \frac{1-{\bar N}^2/{\bar\chi}^2}{1+{\bar \kappa}^2/{\bar\chi}^2} (1+(2{\bar\kappa}^2 + {\bar N}^2)/\bar\chi^2). \label{GGmax} 
\end{equation}

When $a=M$, the asymptotic behaviors of the proper growth rate
${\hat\Gamma}_{max}(r)$ and wavenumber $\bar k_{max}(r)$ for $r \to
r_{ms}$ are 
simply expressed by the asymptotic relation between $\Omega(r)$ and
$\omega(r)$. We show in Fig. 2 the variations of dimensionless
variables, $\bar\chi^2, 
\bar\kappa^2$ and $\bar\Gamma^2_{max}, {\bar k}_{max}$, along the radius
from a event horizon for $a=M, 0$. When $r \to r_{ms}$ and $a
\to M$, the angular velocity of a circular orbit 
coincides with the geometrical angular velocity, $\Omega \to
\omega$. The "Coliori's force" $\kappa^{\hat r} \delta v^{\hat r}$ in
(\ref{eq 
vf}) then reduces to be zero, $\kappa^{\hat r} \delta v^{\hat r} \propto
v_{(\varphi)}\nabla^{\hat r}\ln{(\gamma^2 e^\psi v_{(\varphi)} )} =
(\Omega - \omega)Ar^{-3}\nabla^r \ln{(\gamma^2 e^\psi v_{(\varphi)} )}
\to 0$ for $r \to r_{ms}$. Thus the epicyclic frequency,
$\kappa=(2{\tilde\Omega}_{\hat r}\kappa_{\hat r})^{1/2}$, becomes to be
zero. Since $\bar\chi^2 \ge 2$, the ratio ${\bar N}^2/{\bar \chi}^2$ is
negligibly small in (\ref{GGmax}). The maximum growth rate
${\hat\Gamma}_{max}$ and wavenumber $\bar k_{max}$ at $r \approx    
r_{ms}$ are then simply expressed as,
\begin{equation}
{\hat \Gamma}_{max}^2 \approx  - \frac{\chi^2}{4} \qquad  
{\bar k}_{max}^2 \approx \frac{\bar\chi^2}{4}.
\end{equation}

The value of $\chi^2$ is determined by the strengths of two actions. 
One is the radial acceleration $\delta\dot {v}^{\hat r}$ by the
"centrifugal force", $\delta\dot {v}^{\hat r} \propto
2\tilde\Omega_{\hat r}\delta v^{\hat\varphi}$. 
Other is the induction of magnetic field by the differential rotation of
fluid motion, 
$\delta{\dot B}^{\hat\varphi} \propto \delta B^{\hat r} \gamma^2
v_\Omega \nabla^{\hat r} \ln{\Omega}$. We define the shear term by
${\Omega}_{shear}= \gamma^2 v_{\Omega}\nabla^{\hat
r}\ln({\Omega})$. Then the function $\chi^2$ is expressed as   
${\chi^2} = -2{\tilde\Omega}_{\hat r}{\Omega}_{shear}$. The asymptotic
forms of $\Omega_{shear}$ and ${\tilde\Omega}_{\hat r}$ are represented
as $\Omega_{shear}=-1.5\Omega$ and ${\tilde\Omega}_{\hat r} = \Omega$
for $r \to \infty$, and $\Omega_{shear}= -4\Omega$ and
${\tilde\Omega}_{\hat r} = -2\gamma\nabla^r\omega = (4/\sqrt{3})\omega$
for $r \to r_{ms}$. The efficiencies of the two actions relative to the
angular velocity $\Omega$ or the proper angular velocity
$\hat\Omega$(see Fig. 2) are increased at $r \approx
r_{ms}$. When $r \to r_{ms}$, the function ${\tilde\Omega}_{\hat r}$ in
(\ref{Omeg}) is
represented mainly by the variable related with the differential
rotation of the geometry, $v_\omega \nabla^{\hat r}\ln{\omega}$, since another
variable $v_{(\varphi)}\nabla^{\hat r} \psi$ reduces to be zero, $
v_{(\varphi)}\nabla^{\hat r} \psi = (\Omega - \omega) e^{\psi -
\nu}\nabla^{\hat 
r}\psi = (\Omega - \omega)Ar^{-3}\nabla^r \psi \to 0$. The finite values
of the growth rate ${\hat\Gamma}_{max}$ and of the wavenumber 
${\bar k}_{max}$ at $r=r_{ms}$ are realized by the rotation of the geometry.

The maximum amplitude of the proper growth rate
$|{\hat\Gamma}_{max}(r)|$ increases in the similitude of the Newtonian
angular velocity of a circular orbit $\Omega_{N}=3/4(Mr^{-3})^{1/2}$ for
$r \to r_{ms}$. We show in Fig. 1 the amplitudes of growth rates 
$|{\hat\Gamma}_{max}(r)|$ for $a=M,0$. The ratio 
$|{\hat\Gamma}_{max}(r)| / (3/4(Mr^{-3})^{1/2})$ is restricted to a
finite range as pointed by \citet{G04}. Here its ratio varies between 1
and $ 5/(2\sqrt{3}) \approx 1.44$ when $r \ge r_{ms}$ and $v_A <
0.1c$. The rapidly rotating black hole with $a=M$ produces the rapid
growth rate which is about one-order of magnitude larger than that for
$a=0$,
${\hat\Gamma}_{max}(r_{ms},a=M)=12.5{\hat\Gamma}_{max}(r_{ms},a=0)$. If
the marginally bound orbit $r_{mb}$ or photon orbit $r_{ph}$ is realized
for $a=0$, then the proper growth rate $|{\hat\Gamma}_{max}|$ reaches to
$\infty$. 

It is useful for the magnetohydrodynamics to
evaluate the growth rate in MRI in comparison with the speeds of other
dynamical processes. The angular velocity of a 
circular orbit $\hat\Omega$ is a typical measure to make clear the
characteristic speeds of the growth rates, $\bar\Gamma=\hat\Gamma
/\hat\Omega$, and to make clear the typical wavelenghth of
perturvations, ${\bar k}_{max} = ({\rm\bf k}\cdot{\rm\bf v}_A) /\hat\Omega$. 
We show in Fig. 3 the dispersion relations with use of $\bar\Gamma$ and
${\bar k}$ for $a=M,0$. In the rapidly rotating spacetime with $a=M$, the
relative growth rate $\bar\Gamma$ is observed to be very large in the
vicinity of $r_{ms}$. Its relative value reaches to 
$|\bar\Gamma(r\approx r_{ms})| \approx 3.75 \approx
5 |\bar\Gamma(r=\infty)|$. The unstable range of wavenumber ${\bar
k}_{crit}$ is measured to expand in proportion
to the shear variable ${\bar k}_{crit} \propto \bar\chi$.
On the other hand the relative growth rates 
$\bar\Gamma(r)$ and the critical wavenumbers ${\bar k}_{crit}$ in the
static spacetime $a=0$ little change at the space $r \ge r_{ms}$. 
These characteristic asymmptotic behaiviors of the relative quantities, ${\bar
\Gamma}_{max}(r \to r_{ms})$ and ${\bar k}_{crit}$, are braught about
partially by 
the variations of $\hat\Omega(r \to r_{ms})$. When $a=0$ and $r \to
r_{ms}$, the circular velocity $v_\Omega(r)$ and the proper angular
velocity ${\hat\Omega}(r)$ increase in inverse 
proportion to $(r-r_{h})^n$, while for $a=M$, the circular velocity 
$v_{(\varphi)}(r)$ and the proper angular velocity ${\hat\Omega}(r)$
become constant at $r \approx 
2r_{ms}$. The rapid rotation of the geometry restricts the increases of
$v_{(\varphi)}(r)$ and $\hat\Omega(r)$ for $r \to r_{ms}$ to be
$v_{(\varphi)}=v_\Omega - v_\omega= \infty - \infty \to 1/2$ and
${\hat\Omega} =\gamma \Omega e^{ -\nu} - \gamma\omega e^{
-\nu}=\infty -\infty \to 1/(2\sqrt{3})$. The observers in LNRF(ZAMOs)
evaluate the growth rate $\tilde\Gamma$ to be large in comparison with
the observed angular velocity of a circular motion $\tilde\Omega$.         

Let's consider the dispersion relation over the extended space around a
black hole.  
The rotating black hole distorts the geometry of spacetime not only in
the radial direction but also in the azimuthal direction. The angular
velocity of a circular orbit at a finite height $z$ from the equatorial
plane, $\Omega(r, z)$, was represented by \citet{YUA04}. They assumed the
pressure force in the vertical direction to be balanced with the
gravity. By using its angular velocity, $\Omega(r, z)$, the dispersion
relation on the plane, $z=contant$, is obtained. Here we introduce the cylindrical spatial coordinates
$(\varpi, z, \varphi)$ where $\varpi$ and $z$ are defined by $\varpi =
r\sin{\theta}, z=r\cos{\theta}$. The derivative in $\hat\varpi$ or 
$\hat z$ 
direction is given by the combination of the components of the derivatives in
$\hat r$ and $\hat\theta$ directions, e.g., $\nabla^{\hat \varpi} \Omega =
\sin{\theta} \nabla^{\hat r} \Omega + \cos{\theta}\nabla^{\hat \theta}
\Omega$. The distributions of the maximum amplitude of 
the growth rate, ${\bar\Gamma}_{max}^2(\varpi, z)$, and the wavenumber at the
maximum growth rate, $\bar{k}_{max}(\varpi,z)$, are depicted in Fig.4,
 where the Brunt-V{\rm $\ddot a$}is{\rm $\ddot a$}l{\rm 
$\ddot a$} frequency is assumed to be constant, ${\bar N}_{\hat z}^2 =
N_{\hat z}^2/ {\hat\Omega }^2=0.8$ and $\bar{N}_{\hat
\varpi}^2=0.01{\bar N}_{\hat z}^2$, and the directions of magnetic
 field and wavenumber are fixed to be ${\rm\bf B}=B^{\hat z}{\rm\bf n}_{\hat
 z}$ and ${\rm\bf k}=k_{\hat z}{\rm\bf n}_{\hat z}$.  When $a=M$, the
 amplitude of the maximum growth rate, 
${\bar\Gamma}_{max}^2(\varpi, z)$, is large over the wide area around a black
hole. The wavenumber $\bar{k}_{max}(\varpi,z)$ greatly increases 
near to the event horizon.

\subsection{Magnetorotational Instability exerted by the Magnetic Fields
  Induced by the Differential Rotation of the Orbiting Frame} 

When $\beta \le 1$ and $\gamma^2 v_\Omega v_{(\varphi)}/c^2 \approx 1$,
the term containing $v_A^2 \gamma^2 v_{(\varphi)} v_\Omega$ in the
dispersion relation (\ref{dis}) exerts on the 
MRI. When $k_{\hat r}=0$, the insertion of this term leads to no change
of wave mode in the characteristics of perturbations. The sign of this term
is contradictory to the sign of the change in fluid $N^2$, which
means that this term enhances the 
MRI. The dispersion relation for $a=M$, $r=1.001r_{ms}$, $z=0$ and
$k_{\hat r}=0$ is shown in Fig.5. When Alfven velocity $v_A$ is larger
than $\sim 0.1c$, the growth 
rate is remarkably enhanced by this term and the perturbations with $\bar k =0$ 
become unstable, $\Gamma^2(k=0) < 0$. 

When the wave is propagating in the radial direction,
 $\delta \propto e^{\Gamma_R t + i(\Gamma_It+k_{\hat r}x^{\hat r} +
 k_{\hat \theta} x^{\hat \theta})}$, the waves of perturbations become
 over stable. The unstable branch of the
 dispersion relation in phase space $(k_{\hat\theta}, k_{\hat r})$
 is shown in Fig.7, where the parameters are set to 
 be $v_A=0.5c, \ a=M, \  r=1.001r_{ms}$ and $z=0$. Here the direction of magnetic
 field is fixed to be ${\rm\bf B}=B^{\hat\theta}{\rm\bf n}_{\hat
 \theta}$. The imaginary component of the 
 growth rate $\Gamma_{I}$ is also shown in Fig.7. The wave frequency
 $\Gamma_{I}$ increases in accordance with the radial wavenumber $k_{\hat
 r}$. When the Alfven
 velocity is nonrelativistic, $v_A \ll c$, the 
 unstable branch in the wavenumber space $(k_{\hat\theta}, k_{\hat r})$ is
 restricted in a area $k_{\hat\theta}^2 + k_{\hat r}^2 \le 
 k_{crit}^2$ (see the left pannel in Fig.6). Here $\Gamma_I \approx 0$
 (see the right panel in Fig.6), and then the
 perturbations exponentially grow. When $v_A \ge 0.1c$, the unstable area
 of the phase space $(k_{\hat\theta}, k_{\hat r})$ enlarges to the
 larger radial wave number, $k_{\hat 
 r} > k_{crit}$ (Fig.7). The oscillating waves 
 exponentially increase their amplitudes over
 the wide space of wavenumbers $(k_{\hat \theta}, k_{\hat r})$.

The above instability is caused by the magnetic induction due to the
differential rotation of the orbiting frame, 
\begin{equation}
\frac{\partial{\rm \bf B}}{\partial \hat\tau} \propto -\gamma^2v_{(\varphi)}v_\Omega ({\rm \bf v}\times {\rm\bf B})\times \nabla \ln{(\Omega/{\hat\alpha})}.
\end{equation} 
The perturbation of a radial velocity $\delta v^{\hat r}$ exponentially
increases the magnetic field, 
\begin{equation}
\frac{\partial{\ln{B^{\hat\theta}}}}{\partial \hat\tau} \propto -\gamma^2v_{(\varphi)}v_\Omega {\delta v^{\hat r}} \nabla^{\hat r} \ln{(\Omega/{\hat\alpha})} > 0,\qquad  {\rm when \ \ }  {\delta v^{\hat r}} >0.
\end{equation} 
The increased magnetic field more enhances the motion of fluid according to 
the equations (\ref{eq vr})---(\ref{eq vf}).

\section{Discussion}

\citet{BH92} had shown that the maximum growth rate of MRI agrees with
the local Oort $A$-value of the disk. It is conjectured 
that the Oort $A$-value of a disk is an upper bound to the growth rate
of any instability feeding upon the free energy of differential
rotation. They have investigated the reasons behind this remarkably
general behavior of MRI by using a form of the dynamical Hill
equations(Hill 1878). The Hill equations have found their widest
applicability in the study of collisionless system. We show here that
the maximum growth rate of MRI in curved spacetime is somewhat different
from the general relativistic version analogous to the local Oort $A$-value.

The local Oort $A$-value is the growth rate of the dynamical shear 
instability of a particle motion in the field of central potential. Its
instability is caused by the displacement of the particle orbit in which the
angular velocity of a circular orbit keeps a original value, $\Omega=\Omega_0$. 
When the displacement is positive, i.e. $\delta r >0$, the centrifugal force is then superior to the gravitational force, the radius of the perturbed orbit becomes 
more larger, and vice versa.  

When the motion is restricted in the equatorial plane, 
the geodesic equations in LNRF are represented with the expressions
of the time coordinate, $dx^{(0)}=d\tilde\tau$, and the covariant derivative,
${\tilde\nabla}^{(r)}$, as
\begin{eqnarray}
\frac{d^2 x^{(r)}} {d \tilde{\tau}^2 } &+& \tilde\nabla^{(r)} \ln \alpha + v_\omega  \frac{dx^{(\varphi)}}{d\tilde\tau} \tilde\nabla^{(r)} \ln{\omega} - \Bigl(  \frac{dx^{(\varphi)}}{d\tilde\tau} \Bigr)^2 \tilde\nabla^{(r)} \ln e^\psi = 0, \label{eq: dynr} \\
\frac{d^2 x^{(\varphi)}} {d\tilde{\tau}^2 } &+& \frac{dx^{(\varphi)}}{d\tilde\tau} \frac{dx^{(r)}}{d{\tilde\tau}} \tilde\nabla^{(r)} \ln e^\psi  =0.
\end{eqnarray}
 
Let's consider the situation of perturbations that the particle is allowed to make small excursions from a circular
orbit, $x^{(r)} = e^{\mu_1}R_0, \ dx^{(\varphi)} = e^\varphi(\Omega_0-\omega)dt$, by introducing small
quasi-Cartesian $x$ and $y$ variables:
\begin{equation}
x^{(r)} =e^{\mu_1} R_0+x, \quad\quad  dx^{(\varphi)} = e^\varphi(\Omega_0-\omega)dt + dy. 
\end{equation}

We assume the local force drawing the displaced particle back to its unperturbed
location, $-Kx, \ -Ky$, where $K$ is the some positive constant.   
The linearized equations are written by
\begin{eqnarray}
{\ddot x} - 2 \Omega_{\tilde r} \dot y - \tilde\chi^2 x = -Kx, \qquad 
\ddot y + {\kappa}_{\tilde r} \dot x = -K y. 
\end{eqnarray}
where
\begin{eqnarray}
\Omega_{\tilde r} = v_{(\varphi)} {\tilde\nabla}^{(r)} \ln{e^\psi} - \frac{1}{2} v_\omega  {\tilde\nabla}^{(r)} \ln{\omega}, \quad 
\tilde\chi^2 = - 2 v_\Omega \Omega_{\tilde r} {\tilde\nabla}^{(r)} \ln{\Omega} \ \  {\rm and} \ \ 
{\kappa}_{\tilde r} = 2 v_{(\varphi)} {\tilde\nabla}^{(r)} \ln{e^\psi}¡¢\label{Achi} 
\end{eqnarray}
and the dots denote the time derivatives, $\ddot x=\partial^2 x/\partial {\tilde\tau}^2$.

The $x$ and $y$ displacements have solutions to the above equations varying
as $e^{i{\tilde \Gamma} {\tilde\tau}}$, with
\begin{eqnarray}
{\check\Gamma}^4 - {\check\Gamma}^2({\tilde\kappa}^2 - \tilde{\chi}^2) - K {\tilde\kappa}^2 =0, \label{Adis}
\end{eqnarray}
where ${\check\Gamma}^2 = {\tilde\Gamma}^2 - K$ and ${\tilde\kappa}^2 = 2
\Omega_{\tilde r} \kappa_{\tilde r}$. 
The above dispersion relation leads the following expression of the 
maximum growth rate, 
\begin{equation}
{\tilde\Gamma}^2_{max} = - \frac{{\tilde\chi}^4}{4{\tilde\kappa}^2}. \label{AGmax}
\end{equation}
Far away from a black hole the functions, $\tilde\chi^2$ and $\tilde\kappa^2$,
are approximately represented by 
\begin{eqnarray}
\tilde\chi^2 \sim - \frac{d\Omega^2}{d\ln r}, \quad\quad {\tilde\kappa}^2 \sim \frac{2\Omega}{r}\frac{d(r^2\Omega)}{dr}. 
\end{eqnarray}
Then the maximum growth rate approaches to the local Oort $A$-value:
\begin{eqnarray}
|\tilde\Gamma_{max}| \sim -\frac{1}{2}\frac{d{\Omega}}{d\ln{r}}. 
\end{eqnarray}
Thus the dispersion relation (\ref{Adis}) is the general relativistic
version analogous to the dynamical shear instability in which the
maximum growth rate is the local Oort $A$-value.

We show the variations of the maximum growth rate
${\bar\Gamma}^2_{max}(r)$ and the related functions ${\bar\kappa}^2$
and $\bar\chi^2$ for $a=M, 0$ in Fig.7. Here we represent the 
dimensionless variables defined by ${\bar\Gamma}={\tilde\Gamma}/{\tilde\Omega},\ 
\bar\chi=\chi/{\tilde\Omega},\ {\bar\kappa}
={\tilde\kappa}/{\tilde\Omega}$.

It is a distinctive feature in DSI that at the inner limit, $r\to
r_{ms}$, the maximum growth rate $|{\bar\Gamma}_{max}(r)|$ for $a=M$ 
reaches to $\infty$. Here the function corresponding to the epicyclic
frequency $\tilde\kappa$ approaches to infinitesimal, $\tilde\kappa \to
\sqrt{2}(r-r_{ms}) \to 0$, when $r \to r_{ms}$ while the function
representing the shear $\tilde\chi^2$ is finite at the inner limit,
$\tilde\chi^2 \to 3$. The growth rate thus becomes
$|{\tilde\Gamma}_{max}|=3/(2\sqrt{2}(r/r_{ms}-1))  \to \infty$ for $r\to
r_{ms}$. The maximum growth rate in DSI is 
remarkably different from that in MRI when $a=M$ and $r = r_{ms}$.

This distinctive difference in $\bar\Gamma_{max}$ between MRI and DSI is understood with the dispersion relations, (\ref{Adis}) and (\ref{dis}). The shear term containing the differential angular velocity $\nabla
\ln{\Omega}$, i.e.,  $\tilde\chi^2$, is inserted in the parenthesis at the 
second term on the left hand of (\ref{Adis}) while it is in the third
term in (\ref{dis}). It leads the different expressions in the
maximum growth rate between (\ref{Gmax}) and (\ref{AGmax}). for $r \to r_{ms}$, the maximum growth rate $\bar\Gamma_{max}$ expressed by (\ref{Gmax}) becomes finite when $\bar\kappa^2 \to 0$.

In DSI the term $\tilde\chi^2$ is produced in the perturbation of
``gravitational force'' in the radial direction (\ref{eq: dynr}).      
In MRI the term containg $\nabla
\ln{\Omega}$ appears in the dynamics in the azimuthal direction (\ref{eq
vf}) through the perturbation of the azimuthal component of magnetic
field $\delta B^{\hat\varphi}$. In the Newtonian limit, the angular
velocity with the power law dependence of radius, $\Omega \propto
r^{-3/2}$, reduces the differences of the coefficients in the dispersion
relation between (\ref{dis}) and (\ref{Adis}) to disappear, as follows ;
$\kappa^2 \to \Omega^2,\quad \kappa^2+\chi^2 \to 4\Omega^2$, in
(\ref{dis}) and ${\tilde\kappa}^2-\tilde\chi^2 \to \Omega^2, \quad
{\tilde\kappa}^2 \to 4\Omega^2$ in (\ref{Adis}). However in rotating
spacetime geometry the differential rotation of the geometry, $v_\omega
\nabla\ln\omega$, becomes a main term in these functions. It is concluded that there exist remarkable differences in the dispersion relation between MRI and DSI.

The behavior of maximum growth rate $\tilde\Gamma_{max}(r)$ in DSI 
is fairly different from that given by 
\citet{G04}. He showed that the maximum growth rate normalized by Newtonian
Oort $A$-value, $|\Gamma_{max}|/[(3/4)(Mr^{-3})^{-1/2}]$, is restricted between 1
and $4/3$ for all radii and for all $a$. His method deriving the
dispersion relation is different in some aspects from ours. He had analyzed 
the dynamics of perturbations in the Boyer-Lindquist coordinate frames (BLCF) in which
the observer measures the motion of a particle with clock and rule at
the distance far from a hole. We have also re-derived the dispersion
relation of DSI in BLCR with the similar forms as in LNRF, which is
shown in Appendix B. Then the function corresponding to the epicyclic
frequency $\tilde\kappa$ reaches to $\infty$ when $r \to r_{ms}$ and
$a \to 1$. Thus the maximum growth rate at $r \to r_{ms}$ becomes infinitesimal, $\Gamma_{max}
\to 0$, which is contrary to the above result in LNRF. 

When $a=M$, the asymptotic behavior of $\kappa_{\tilde r}$ at $r \to
r_{ms}$ is extremely different between BLCF and LNRF. This difference is
caused by the expression of the circular velocity measured in each of
frame. In BLCF the coefficient $\kappa_{\tilde r}$  in (\ref{Akappa}) is described with the velocities, $v_\Omega$
and $v_\omega$ which become infinity at $r \to r_{ms}$, while in LNRF $\kappa_{\tilde
r}$ in (\ref{Achi}) is expressed with $v_{(\varphi)}$ which is limited within
$1/2$ for $r \le r_{ms}$. The function $\kappa_{\tilde r}$ in
(\ref{Achi}) (LNRF) reduces to zero for $r \to r_{ms}$ since the
covariant derivative ${\tilde\nabla}^{(r)} \ln{e^\psi}$ becomes infinitesimal at 
$r \to r_{ms}$. The dynamics in the azimuthal direction is exerted by
the "Coriolis force" $\tilde\kappa_{\tilde r} \dot x$. In BLCF the
rapidly rotating particle perturbed by $\dot x$ is extremely accelerated
in the azimuthal direction. The dynamics of perturbed motion at $r \sim
r_{ms}$ is very different between BLCF and LNRF. 

The perturbed quantities in LNRF are not same as in BLCF. The perturbed velocity in the radial direction in LNRF, $\dot x$, is divided into two components, radial
and azimuthal ones, in BLCF. Then its azimuthal component becomes not always a
first order of small quantity in BLCF, and vice versa. Thus the dispersion relation becomes definitely different between LNRF and BLCF.

The LNRF is suitable for the analysis of perturbations around a rotating black hole since the dragging of inertial frames becomes so severe in BLCF that the $t$ coordinate basis vector $(\partial/\partial t)$ goes spacelike at the static limit $r_0$. Thus it is concluded that the both types of instabilities, DSI and MRI, are remarkably enhanced in the vicinity of a rapidly rotating black hole, though their growth rates are not same.

A summary of the results is as follows. 
The dispersion relation in MRI is represented in general relativistic
forms. The growth rate at $r \approx r_{ms}$ is realized for $a=M$ by the rotation of
spacetime geometry. The maximum growth rate $|{\bar\Gamma}_{max}|$ at $r=r_{ms}$
for $a=M$ is about one-order of magnitude larger than that for $a=0$. When the magnetic field is so intensive that $v_A \ge 0.1c$, the over stability mode becomes remarkable and its growth rate increases according to the strength of magnetic field. The shear instability, MRI or DSI, is largely enhanced by the rapidly rotating black hole though the value of maximum growth rate in DSI is fairly different from that in MRI. The dispersion relation at $r \approx r_{ms}$ is suitably evaluated in LNRF. In the static geometry the growth rates in MRI and DSI remain almost in similar variation as in Newtonian case.

\acknowledgments

We are grateful to Professors T. Ishizuka and T. Yoshida for useful discussion and advice on the magnetorotational instability in curved spacetime.

\appendix
\section{Coefficients of the Affine Connection in the Orbiting Frame}

The rotation one-forms $\omega^{\mu}_{\nu}$ are derived from the
one-forms ${\vec \omega}^{\lambda}$ shown in (\ref{eq: one-form})
\citep{P01,MTW}.   
From the one-form $\vec\omega^{\nu}=\beta^\nu_\lambda {\rm\bf
d}x^\lambda$, we compute the exterior derivative
\begin{equation}
{\rm\bf d}\vec\omega^{\nu}=\frac{\partial \beta^\nu_\lambda}{\partial x^\mu}
 {\rm\bf d}x^\mu \wedge {\rm\bf d}x^\lambda = -\omega^\nu_\mu \wedge \vec\omega^{\mu}.
\end{equation}
The coefficients of Affine connection $\Gamma^{\nu}_{\mu \  \lambda}$
are read off the representation ${\omega}^{\nu}_{\mu} =
\Gamma^{\nu}_{\mu \  \lambda}  \  {\vec \omega}^{\lambda}$. 

Their explicite expressions are as follows:
\begin{eqnarray}
&\Gamma^{\hat t}_{\hat r \hat t} &= e^b \nabla^{\hat r}\ln{\hat\alpha}
- v_{(\varphi)}v_\Omega \gamma^2 \nabla^{\hat r} \ln(\gamma e^\psi v_{(\varphi)}),
\nonumber \\
&\Gamma^{\hat t}_{\hat r \hat \varphi} &=  \frac{1}{2}{\gamma}^2 (
v_\Omega \nabla^{\hat r} \ln\Omega - e^b v_{(\varphi)} \nabla^{\hat r}
\ln(\gamma e^\psi v_{(\varphi)}/{\hat\alpha}),
\nonumber \\
&\Gamma^{\hat t}_{\hat\theta  \hat t} &= e^b\nabla^{\hat\theta}
\ln{\hat\alpha} - v_{(\varphi)}v_\Omega\gamma^2 \nabla^{\hat \theta} \ln(\gamma e^\psi v_{(\varphi)}) ),
\nonumber \\
&\Gamma^{\hat t}_{\hat\theta  \hat \varphi} &=  \frac{1}{2}
v_\Omega\gamma^2\nabla^{\hat\theta} \ln\Omega - e^b v_{(\varphi)}
\nabla^{\hat\theta} \ln(\gamma e^\psi v_{(\varphi)}/{\hat\alpha}), 
\nonumber \\
&\Gamma^{\hat t}_{\hat \varphi \hat r} &=\Gamma^{\hat
r}_{\hat \varphi \hat t} = \frac{1}{2} v_\Omega \gamma^2
\nabla^{\hat r} \ln \Omega,
\nonumber \\
&\Gamma^{\hat t}_{\hat \varphi \hat\theta} &=\Gamma^{\hat
\theta}_{\hat \varphi \hat t} = \frac{1}{2} v_\Omega \gamma^2 \nabla^{\hat\theta} \ln \Omega,
\nonumber \\
&\Gamma^{\hat r}_{\hat \theta \hat r} &= \nabla^{\hat \theta}\ln e^{\mu_1}, \quad \Gamma^{\hat r}_{\hat\theta \hat\theta} = -\nabla^{\hat r}\ln e^{\mu_2},
\nonumber \\
&\Gamma^{\hat r}_{\hat \varphi \hat \varphi} &= -\nabla^{\hat r}
\ln(\gamma e^\psi) + v_{(\varphi)}v_\Omega \gamma^2 \nabla^{\hat r} \ln \Omega,
\nonumber \\
&\Gamma^{\hat\theta}_{\hat \varphi \hat \varphi} &= -\nabla^{\hat
\theta}\ln(\gamma e^\psi) + v_{(\varphi)}v_\Omega \gamma^2
\nabla^{\hat \theta} \ln \Omega. \nonumber
\end{eqnarray}

\section{The Dynamical Shear Instability in the Boyer-Lindquist Coordinate Frame}

The equation of motion of a particle is described in the Boyer-Lindquist
coordinate frame(BLCF) by,
\begin{equation}
\frac{d^2x^\lambda}{d\tau^2} + \Gamma^\lambda_{\mu \nu} \frac{dx^\mu}{d\tau} \frac{dx^\nu}{d\tau} = 0,
\end{equation}
where $\Gamma^\lambda_{\mu \nu}$ are the coefficients of Affine connection in BLCF.
When the motion is restricted in the equatorial plane, 
these equations are represented by using the expressions of coordinates,
$x^{(r)}, x^{(\varphi)}$, and the covariant derivative,
${\tilde\nabla}^{(r)}$, in LNRF as
\begin{eqnarray}
\frac{d^2x^{(r)}} {\alpha^2 dt^2} &+& \tilde\nabla^{(r)} \ln \alpha + v_\omega \frac{dx^{(\varphi)}}{\alpha dt} \tilde\nabla^{(r)} \ln (\omega e^\psi) -  \Bigl(\frac{dx^{(\varphi)}}{\alpha dt}\Bigr)^2 \tilde\nabla^{(r)} \ln e^\psi = 0, \\
\frac{d^2x^{(\varphi)}} {\alpha^2 dt^2} &+& \Bigl\{ v_\Omega \Bigl[2 \tilde\nabla^{(r)} \ln e^{\varphi} +  v_\omega^2 \tilde\nabla^{(r)} \ln \omega  \Bigr]              
 \nonumber \\
 &-& v_\omega \Bigl[ 2 \tilde\nabla^{(r)} \ln v_\omega - (1-v_\omega^2) \tilde\nabla^{(r)} \ln \omega \Bigr] \Bigr\} \frac{d x^{(r)}}{\alpha dt} =0.
\end{eqnarray}
 
Let's consider the situation of perturbations that the particle is allowed to make small excursions from a circular
orbit, $r=R_0, \ \varphi = \Omega_0 t$, by introducing small
quasi-Cartesian $x$ and $y$ variables:
\begin{equation}
r = R_0+x e^{-\mu_1}, \quad\quad  \varphi= \Omega_0t + y e^{-\psi}. 
\end{equation}
The linearized equations are then written by
\begin{eqnarray}
{\ddot x} &-& 2v_{(\varphi)}{\dot y} \tilde\nabla^{(r)} \ln e^\psi + {\dot y}v_\omega \tilde\nabla^{(r)} \ln (\omega e^\psi) \nonumber \\
&+& e^{2(\psi -\nu)} \tilde\nabla^{(r)}\Omega^2 \Bigl[\nabla^{(r)} \ln e^\psi - \frac{\omega}{2\Omega}  \tilde\nabla^{(r)} \ln(\omega e^{3\psi})  \Bigr] x = -Kx, \\
\ddot y &+& \Bigl\{ v_\Omega \Bigl[2\tilde\nabla^{(r)} \ln e^{\varphi}+ v_\omega^2 \tilde\nabla^{(r)} \ln \omega  \Bigr]   \nonumber \\
 &-&  v_\omega \Bigl[2 \tilde\nabla^{(r)} \ln v_\omega - (1-v_\omega^2) \tilde\nabla^{(r)} \ln \omega  \Bigr]\Bigr\} \dot x = -Ky,
\end{eqnarray}
We follow the convention of writing dots for time derivatives, e.g., $\ddot x
= d^2x/(\alpha^2 dt^2) = d^2x/d{\tilde\tau}^2$.  

The $x$ and $y$ displacements have solutions to the above equations varying
as $e^{i{\tilde \Gamma} {\tilde\tau}}$, with
\begin{eqnarray}
{\tilde\Gamma}^4 - {\tilde\Gamma}^2({\tilde\kappa}^2 - \tilde\chi^2 + 2K) + K(K - {\tilde\chi}^2) =0, 
\end{eqnarray}
where
\begin{eqnarray}
\tilde\chi^2 &=& -e^{2(\psi -\nu)} \tilde\nabla^{(r)}\Omega^2 \Bigl[\tilde\nabla^{(r)} \ln e^\psi - \frac{\omega}{2\Omega} \tilde\nabla^{(r)} \ln(\omega e^{3\psi})  \Bigr],\\
{\tilde\kappa}^2 &=& 4v_\Omega v_{(\varphi)} \Bigl\{ \Bigl[ \tilde\nabla^{(r)} \ln e^{\varphi}+ \frac{1}{2} v_\omega^2 \tilde\nabla^{(r)} \ln \omega  \Bigr] - \frac{\omega}{\Omega} \Bigl[ \tilde\nabla^{(r)} \ln v_\omega - \frac{1}{2}(1-v_\omega^2) \tilde\nabla^{(r)} \ln \omega  \Bigr] \Bigr\} \nonumber \\
&\times& \Bigl[\tilde\nabla^{(r)} \ln e^\psi - \frac{\omega}{2(\Omega -\omega)} \tilde\nabla^{(r)} \ln(\omega e^{\psi})  \Bigr]. \label{Akappa}
\end{eqnarray}

The above dispersion relation gives the
maximum growth rate, 
\begin{equation}
{\tilde\Gamma}^2_{max} = - \frac{{\tilde\chi}^4}{4{\tilde\kappa}^2}. \label{AGmaxA}
\end{equation}
Far away from a black hole the functions, $\tilde\chi^2$ and $\tilde\kappa^2$,
are approximately represented by 
\begin{eqnarray}
\tilde\chi^2 \sim - \frac{d\Omega^2}{d\ln r}, \quad\quad {\tilde\kappa}^2 \sim \frac{2\Omega}{r}\frac{d(r^2\Omega)}{dr}. 
\end{eqnarray}
The above maximum growth rate then has the local Oort $A$-value:
\begin{eqnarray}
|\tilde\Gamma_{max}| \sim -  \frac{1}{2} \frac{d{\Omega}}{d\ln{r}}. 
\end{eqnarray}

We show in Fig.9 the maximum growth rate  $\bar{\Gamma}_{max}^2(r)$ and the
related functions, 
${\bar{\kappa}}^2(r), \ \bar\chi^2(r)$ for $a=M$. Here we use the angular velocity of a circularly orbiting particle measured
in LNRF, ${\tilde\Omega}\equiv (\Omega - \omega)/\alpha$, to define the
dimentionless variables such as
${\bar\Gamma}={\tilde\Gamma}/{\tilde\Omega},
\bar\chi=\chi/{\tilde\Omega}$ and $
{\bar\kappa}={\tilde\kappa}/{\tilde\Omega}$.  The absolute
value of the growth rate, $|\bar{\Gamma}_{max}|$, 
for $a=M$ has an upper bound, $|\bar{\Gamma}_{max}| \le 1.5$, which is
a contrast to the maximum growth rate in LNRF. When $a=0$, there is no large
variation in $\bar{\Gamma}_{max}^2(r)$ for the space, $r \ge r_{ms}$. 

At the inner limit, $r\to r_{ms}$, the corresponding dimensionless functions for
$a=M$ become
\begin{eqnarray}
\bar\chi^2 \to 3\cdot 2^8, \quad\quad {\bar\kappa}^2 \to \infty. 
\end{eqnarray}
The inifinity of $\bar\kappa^2$ is brought about by $v_\omega \to
\infty$ when $r\to r_{ms}$. 
However the corresponding function ${\bar\kappa}^2$ 
in the case of LNRF reduces to infinitesimal at $r \to r_{ms}$. While the
growth rate expressed by (\ref{AGmaxA}) reduces to infinitesimal 
at $r \to r_{ms}$, the growth rate expressed by (\ref{AGmax}) reaches
to $\infty$. In the curved spacetime the dynamical shear instability in
BLCF has the form unlike as in LNRF.

\clearpage

\begin{figure}
\figurenum{1}
\plottwo{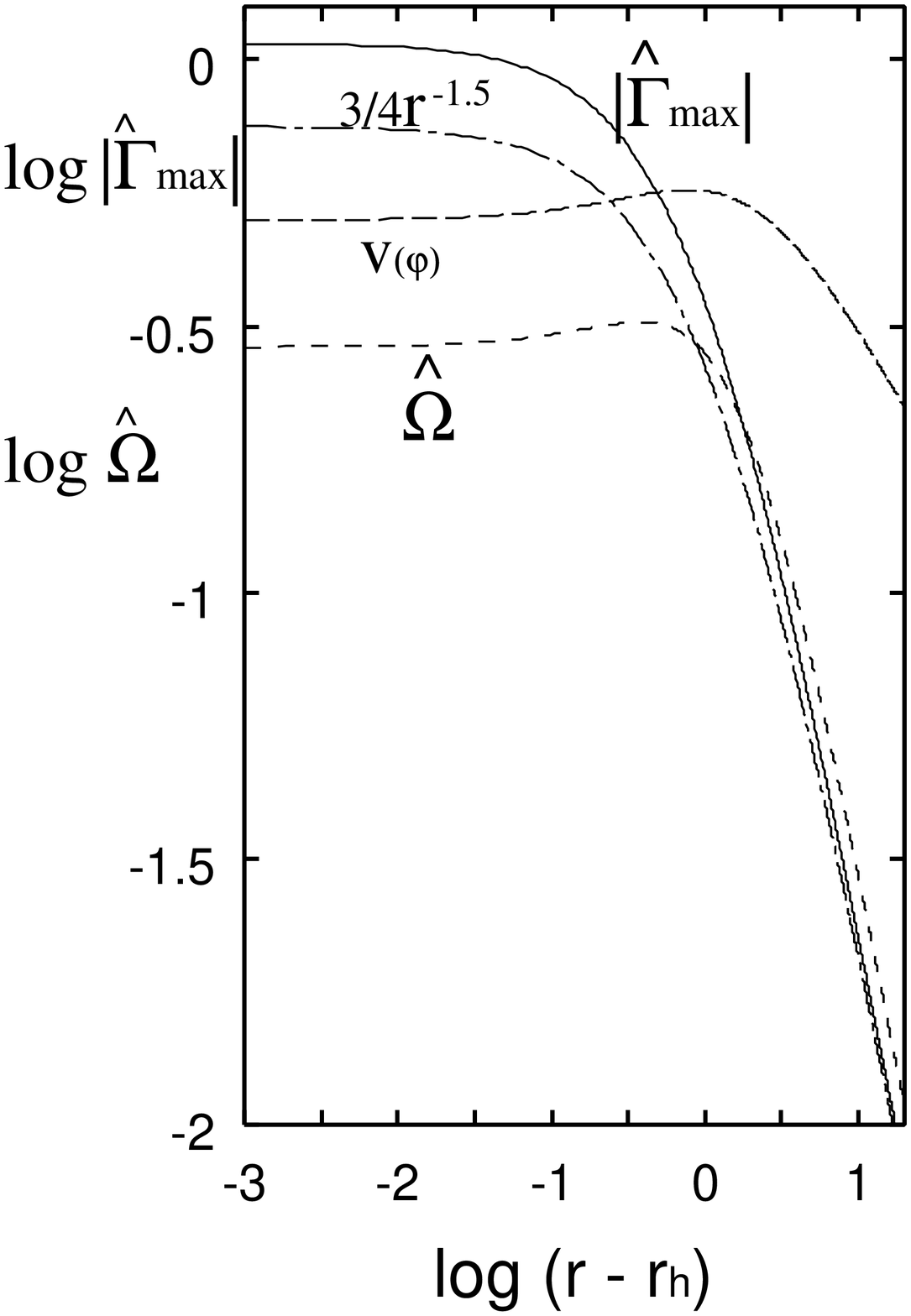}{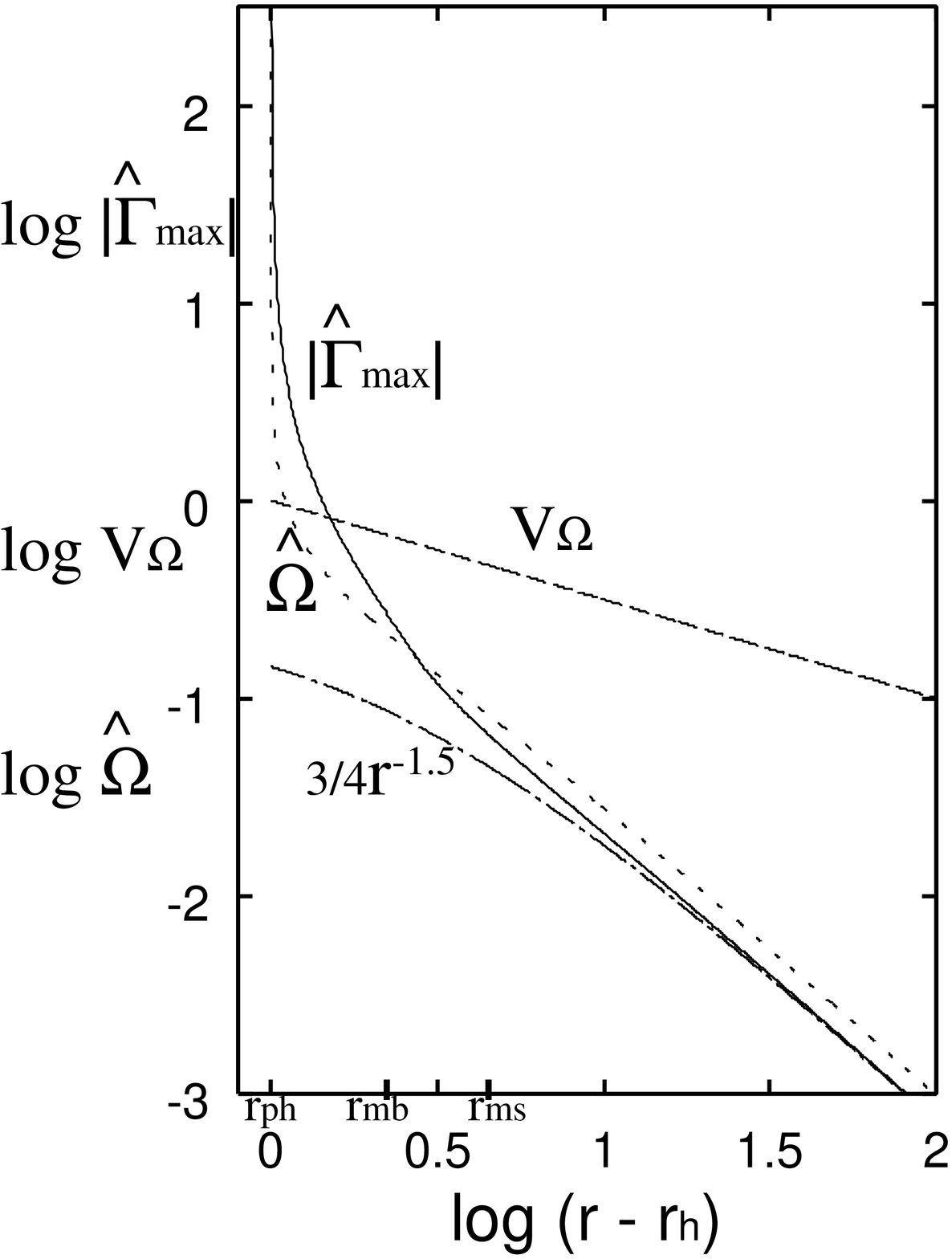}
\caption{ The distributions of the maximum proper growth rate
 $|{\hat\Gamma}_{max}(r)|$, proper angular velocity ${\hat\Omega}(r)$
 and the function $3/4 r^{-3/2}$  
on the distance from the event horizon. The left figure is shown in the
 case of the extreme Kerr metric and the right one is that of
 Schwarzscild metric. Other parameters are $k^{\hat r}=0, N_{\hat\theta}^2=0.8 {\hat\Omega}^2, N_{\hat r}^2=0.01N_{\hat\theta}^2$. 
\label{fig1}}
\end{figure}

\begin{figure}
\figurenum{2}
\plottwo{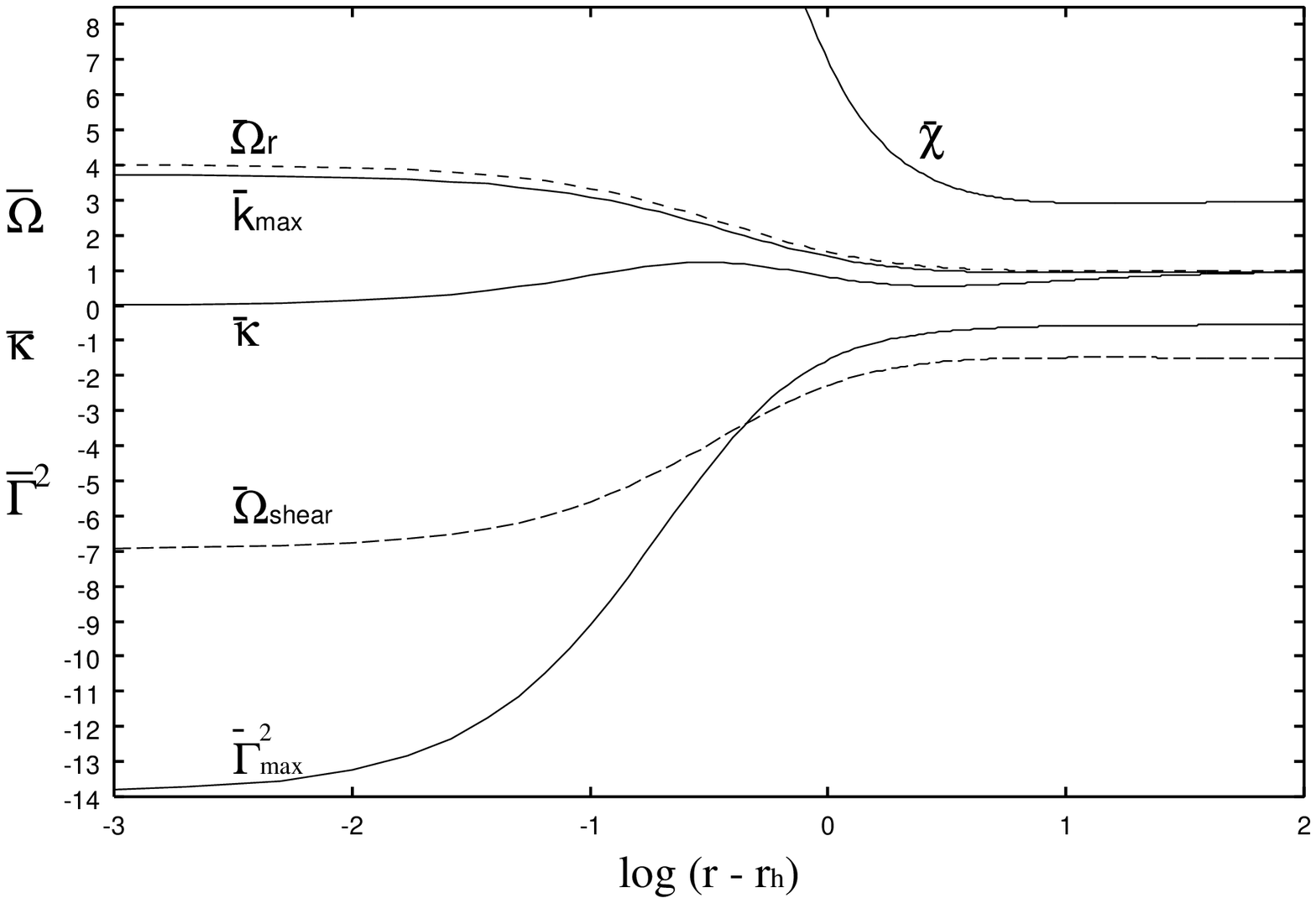}{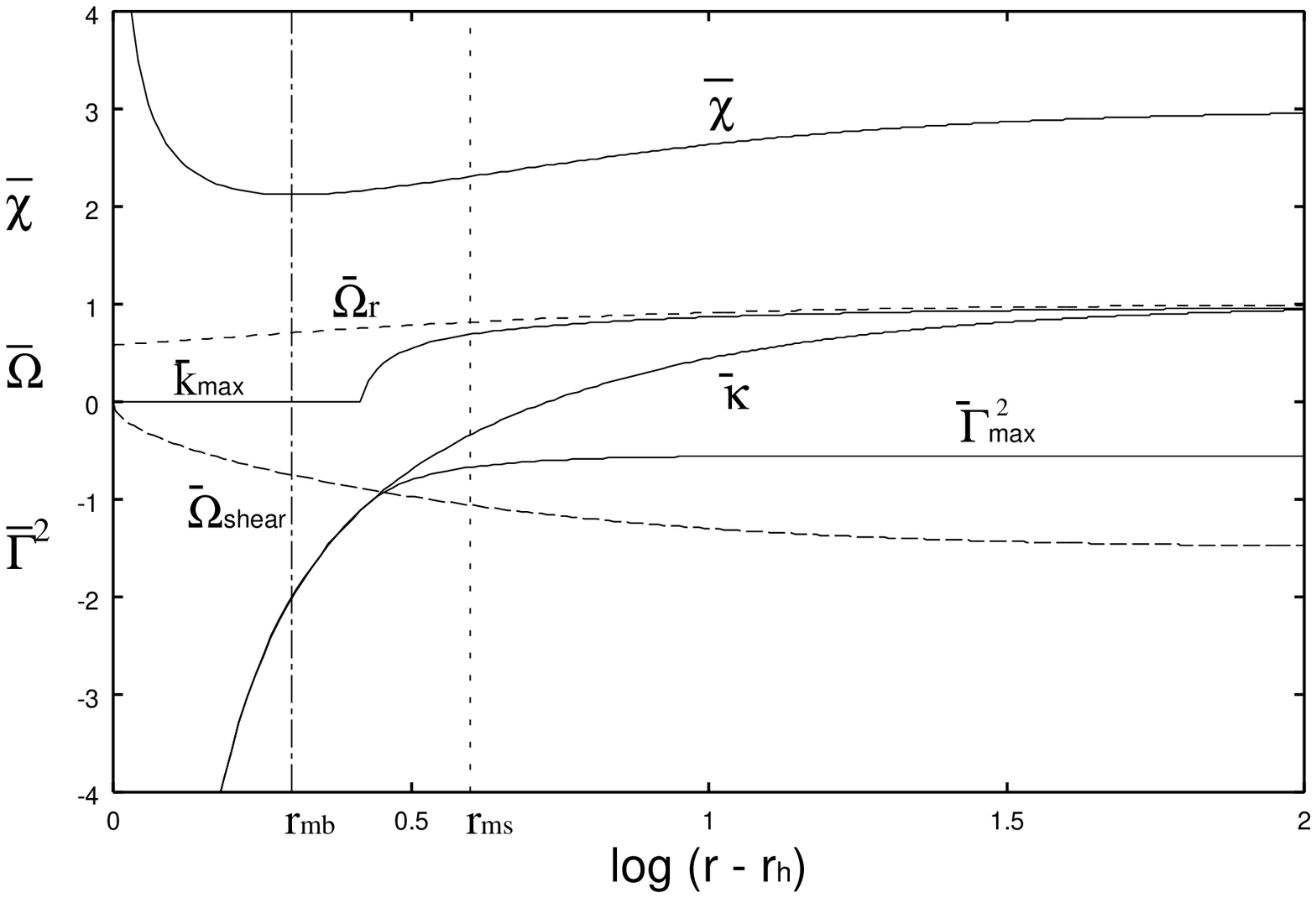}
\caption{The distributions of the maximum relative growth rate ${\bar\Gamma}^2_{max}(r)$ and
 the related terms ${\bar k}_{max}(r), {\bar\chi}^2(r), {\bar\Omega}_{\hat r}(r)$ and ${\bar\Omega}_{shear}(r)$ on the distance from the event horizon: The left figure shows the case of the extreme Kerr metric. The right figure shows the case of the Schwarzscild metric. Other parameters are same as in
 Fig.1. 
\label{fig2}}
\end{figure}

\begin{figure}
\figurenum{3}
\plottwo{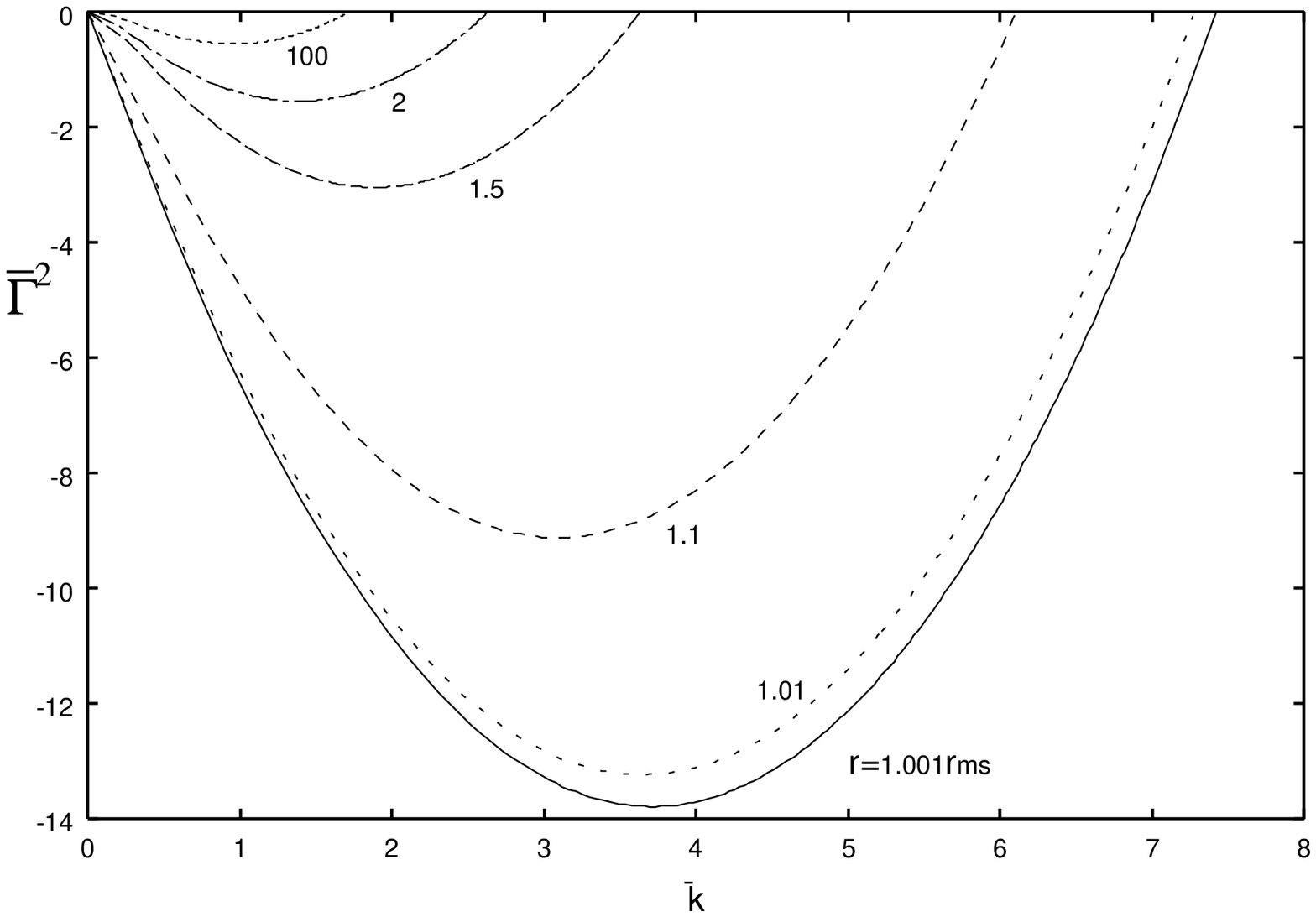}{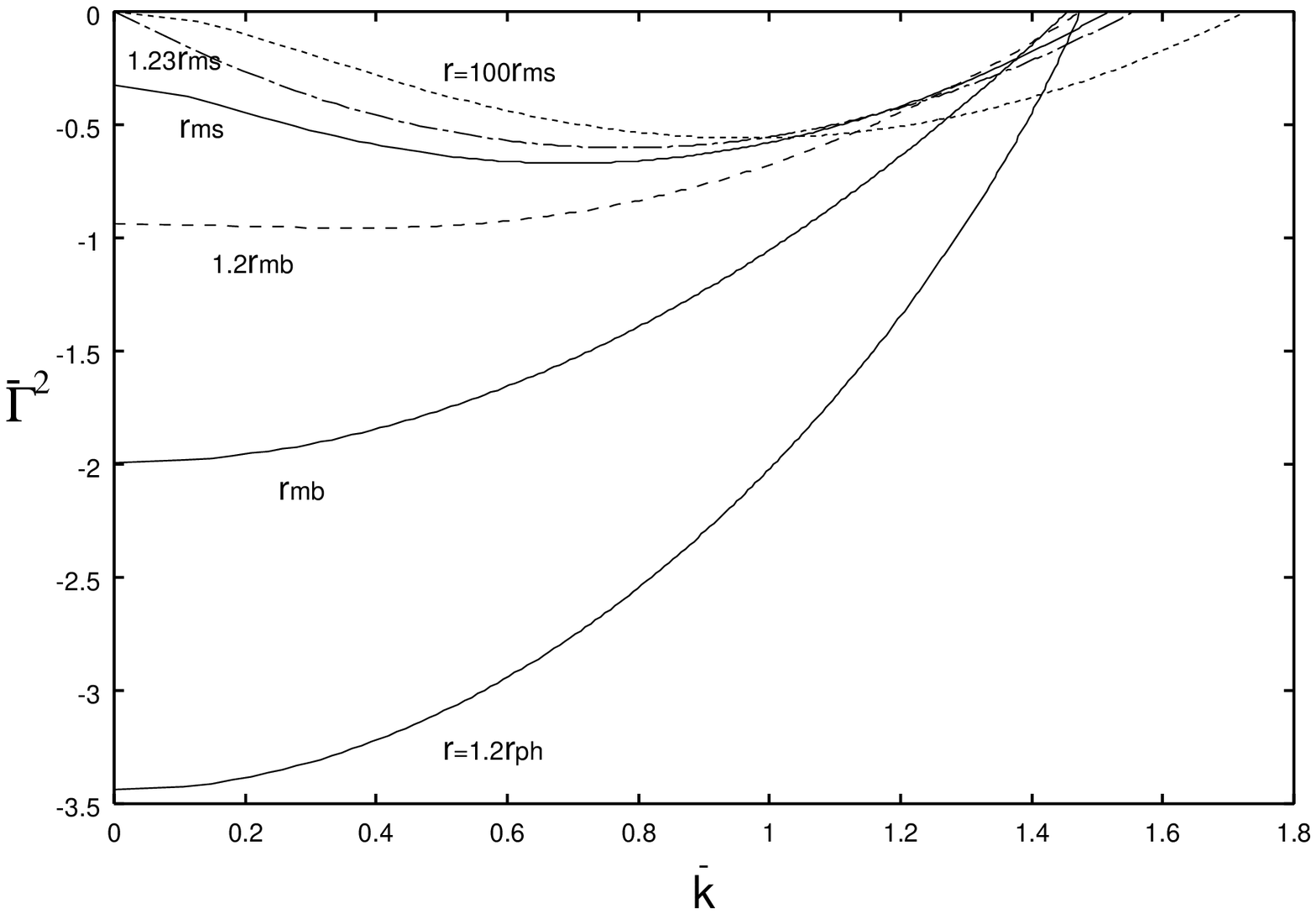}
\caption{Dispersion relation: The Left figure shows the relative growth
 rates for the metric of an extreme Kerr. The squares of growth rate
 relative to the angular  
velocity ${\bar\Gamma}^2=(\hat\Gamma/{\hat\Omega})^2$. are depicted. The
 wavenumber is normalized as $\bar k = {\rm\bf k}\cdot{\rm\bf v}_A/{\hat\Omega}$. The curves are correspond in order to $r=10^2, 2, 1.5, 1.1, 1.01, 1.001(r_{ms})$ from the top. The right figure shows the relative growth rates for the Schwarzscild metric. The curves are correspondent to $r=10^2r_{ms}, 1.23r_{ms}, 1r_{ms}, 1.2r_{mb}, 1.0r_{mb}, 1.2r_{ph}$. Other parameters are same as in Fig. 1. 
\label{fig3}}
\end{figure}

\begin{figure}
\figurenum{4}
\plottwo{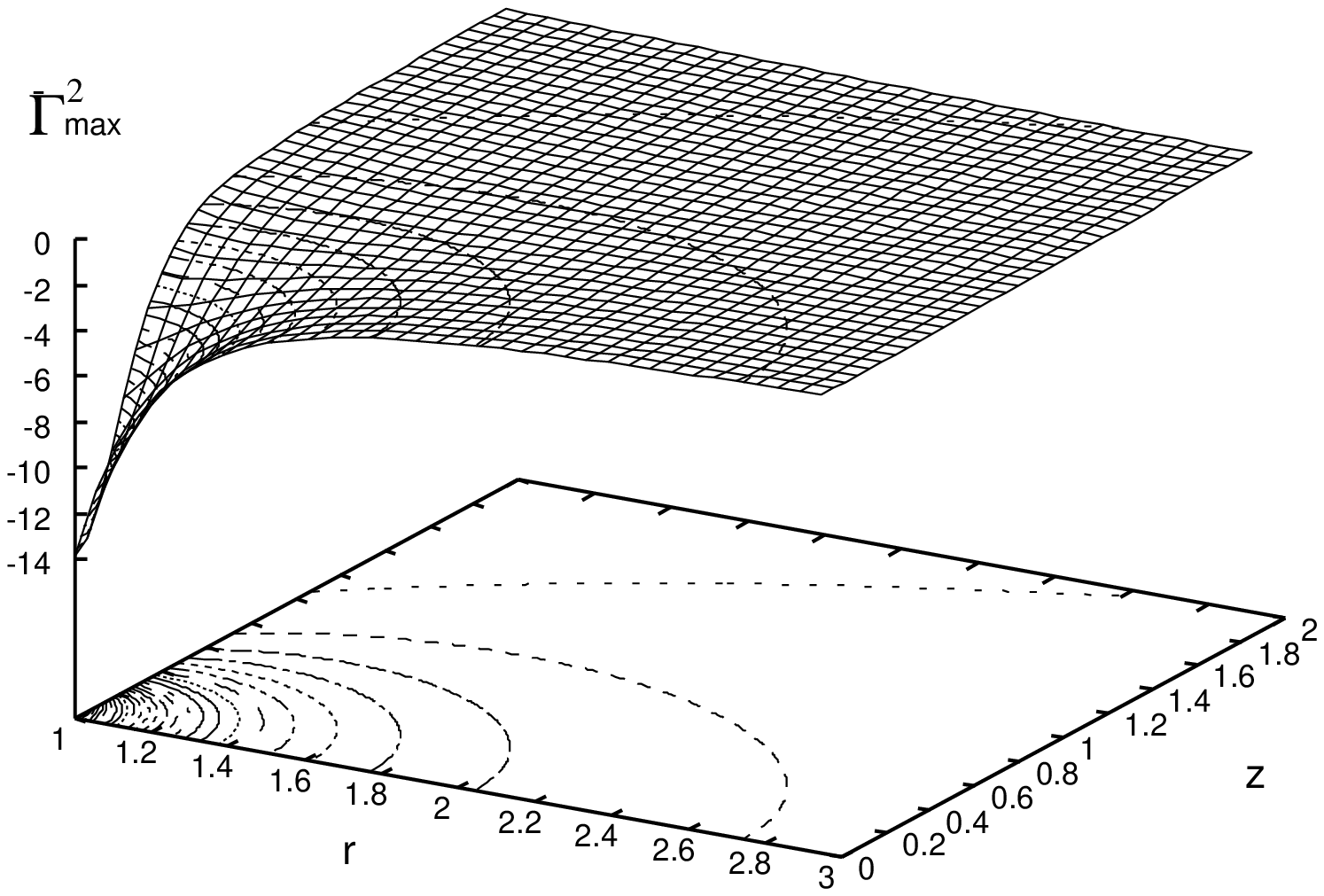}{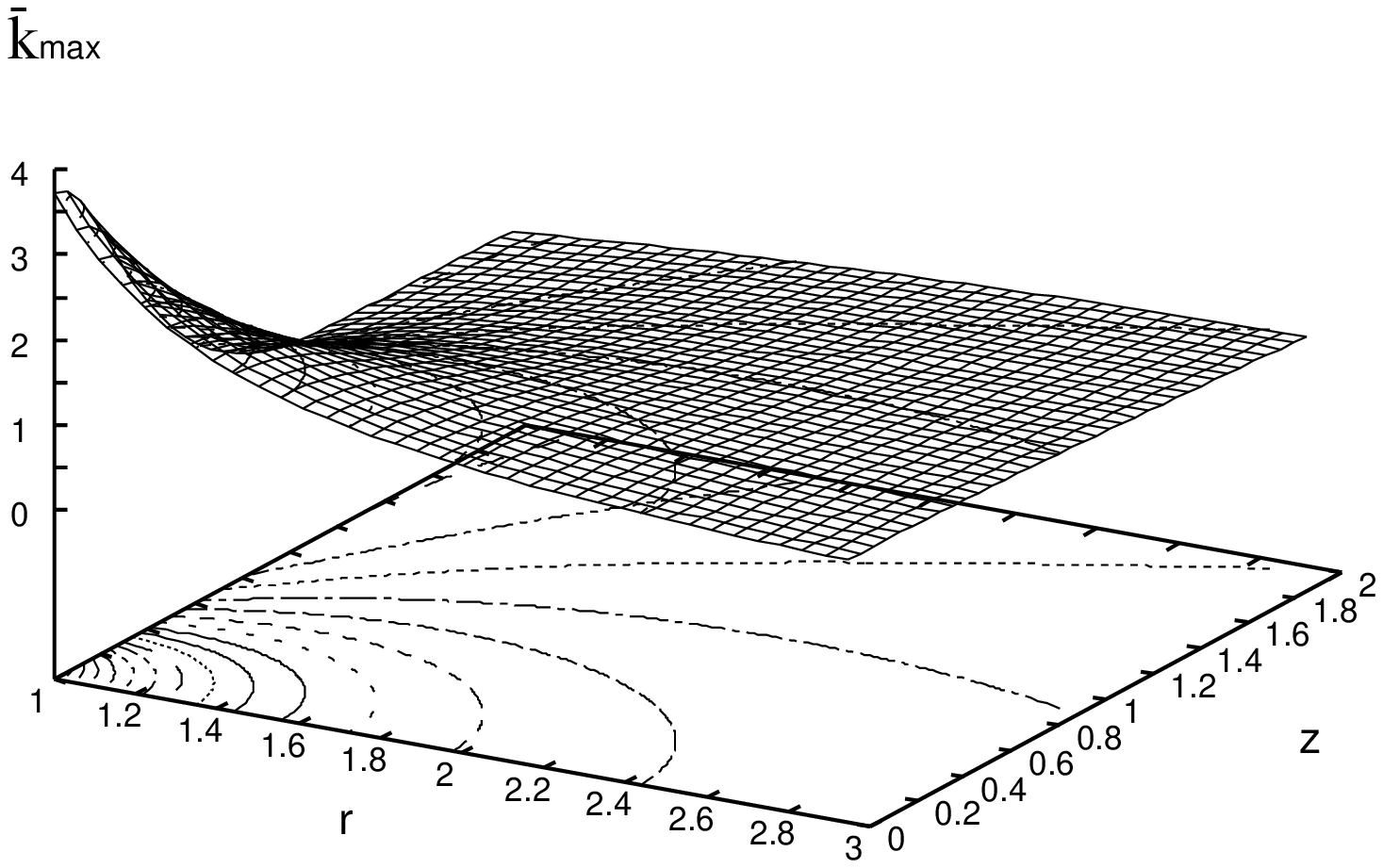}
\caption{The left figure shows the distribution of the maximum relative
 growth rate 
 ${\bar\Gamma}_{max}^2(r,z)$ for $a=M$. 
The right figure shows the distribution of wave number
 ${\bar{k}}_{max}(r,z)$ at the 
 maximum growth rate. Other parameters are same as in Fig. 1. 
\label{fig4}}
\end{figure}

\begin{figure}
\figurenum{5}
\plotone{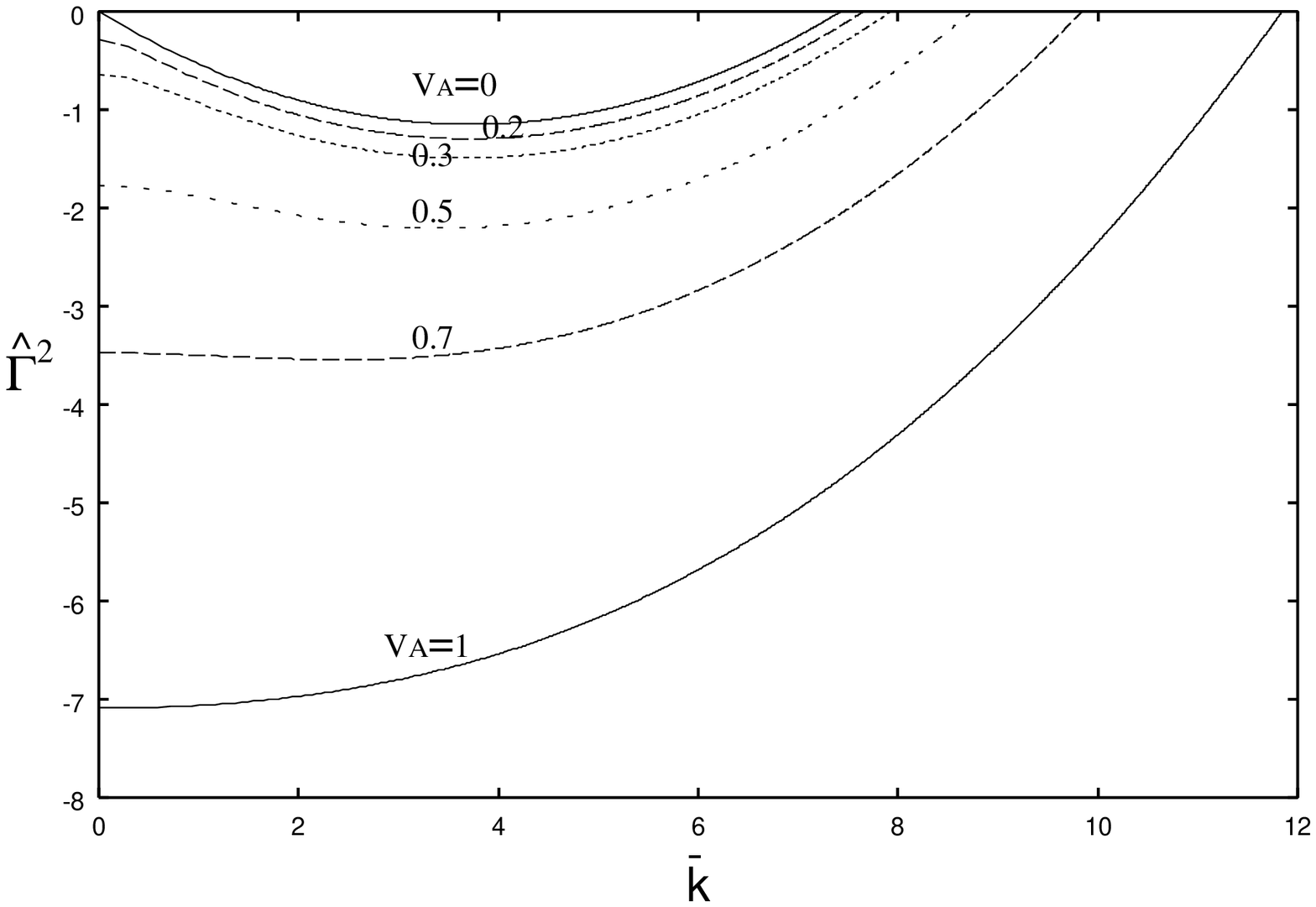}
\caption{Dispersion relations with the term of magnetic induction due to
 the differential rotation of the orbiting frame. The proper growth
 rates $\hat{\Gamma}^2$ are shown as the functions of the normalized
 wavenumber $\bar k = {\rm\bf k}\cdot{\rm\bf v}_A/{\hat\Omega}$. The
 curves are 
 corresponding to the values of Alfven velocity in unit of light
 velocity. The metric of spacetime is an extreme Kerr. The position is
 at $r=1.001r_{ms}$. Other parameters are same as in Fig. 1.
\label{fig5}}
\end{figure}

\begin{figure}
\figurenum{6}
\plottwo{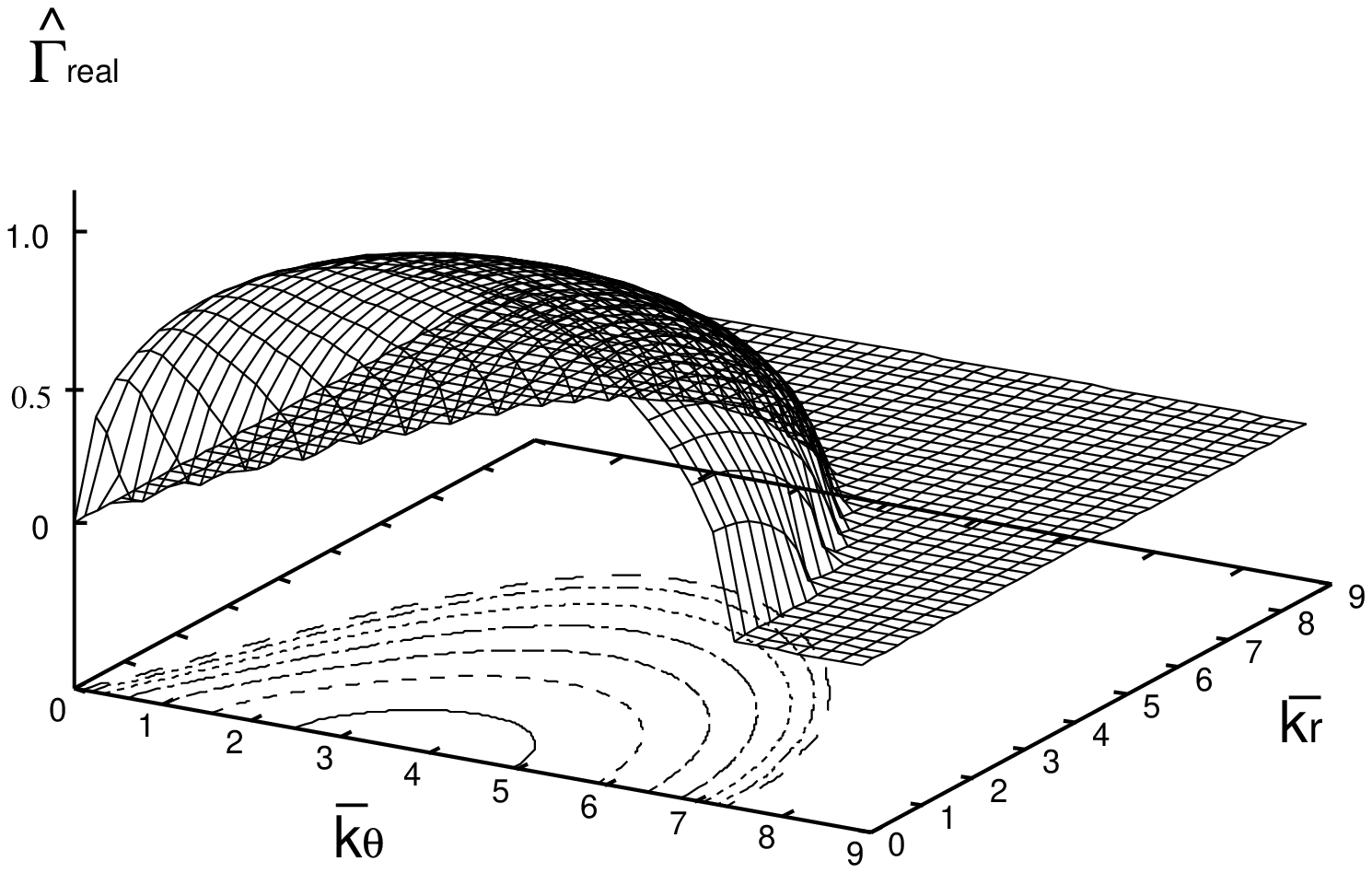}{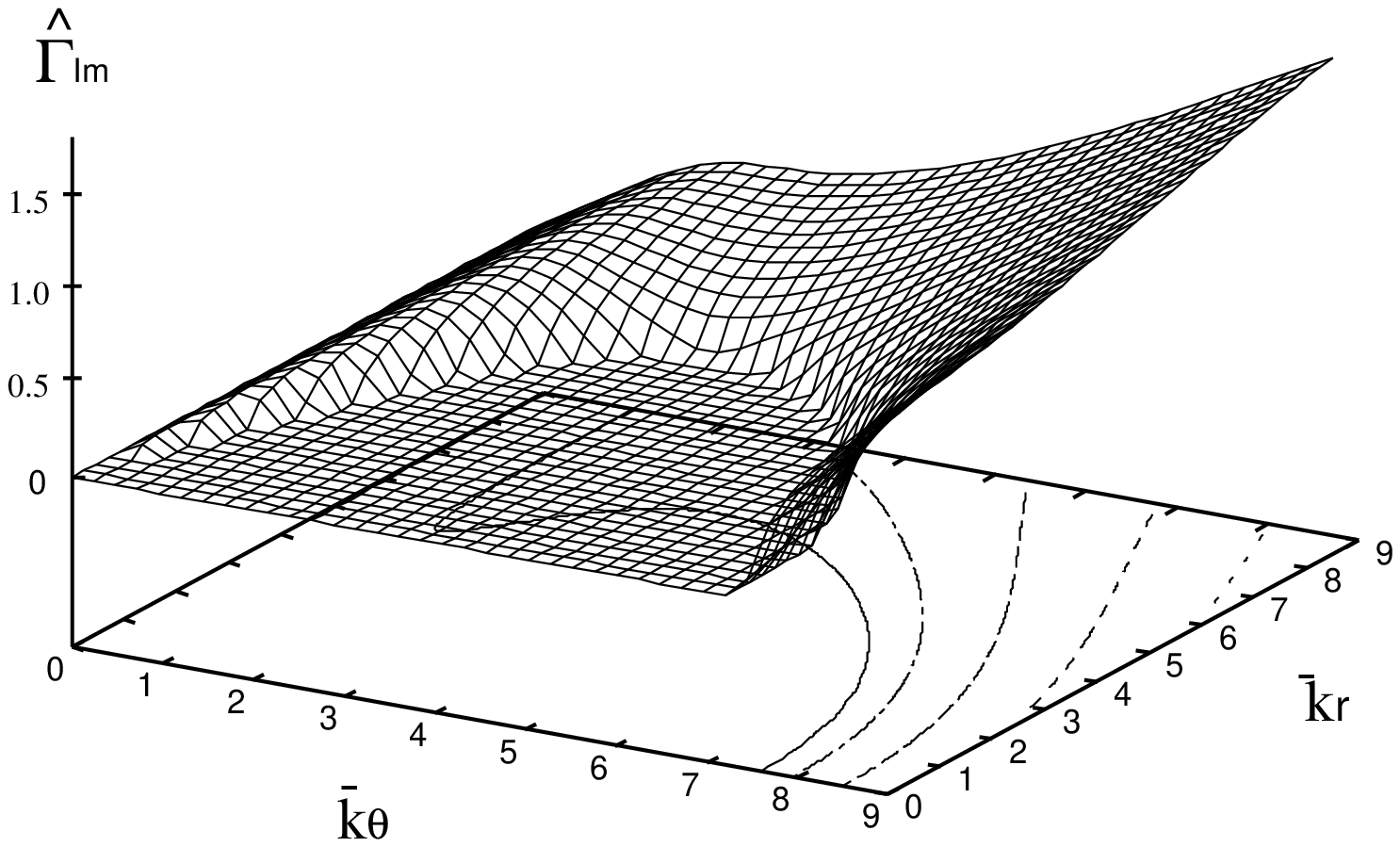}
\caption{Three-dimensional plot of the unstable branch of the dispersion
 relation with $v_A=10^{-3}$. Left figure: The real component of the growth rate
 ${\bar\Gamma}_{real}(k_{\hat\theta}, k_{\hat r})$ is plotted in the
 ${\bar k}_\theta ( = k_{\hat{\theta}} v_A / {\hat\Omega} ), {\bar k}_r( = k_{\hat{r}} v_A / {\hat\Omega})$ plane. The direction of magnetic field is fixed to be ${\rm\bf
 B}=B^{\hat\theta}{\rm\bf n}_{\hat \theta}$. The spacetime of the extreme
 Kerr is selected and the position is set to be 
 $r=1.001r_{ms}$. Other parameters are same in Fig. 1. 
 Right figure: Two-dimensional plot of the imaginary component of the growth
 rate  ${\bar\Gamma}_{Im}(k_{\hat\theta}, k_{\hat r})$. 
\label{fig6}}
\end{figure}

\begin{figure}
\figurenum{7}
\plottwo{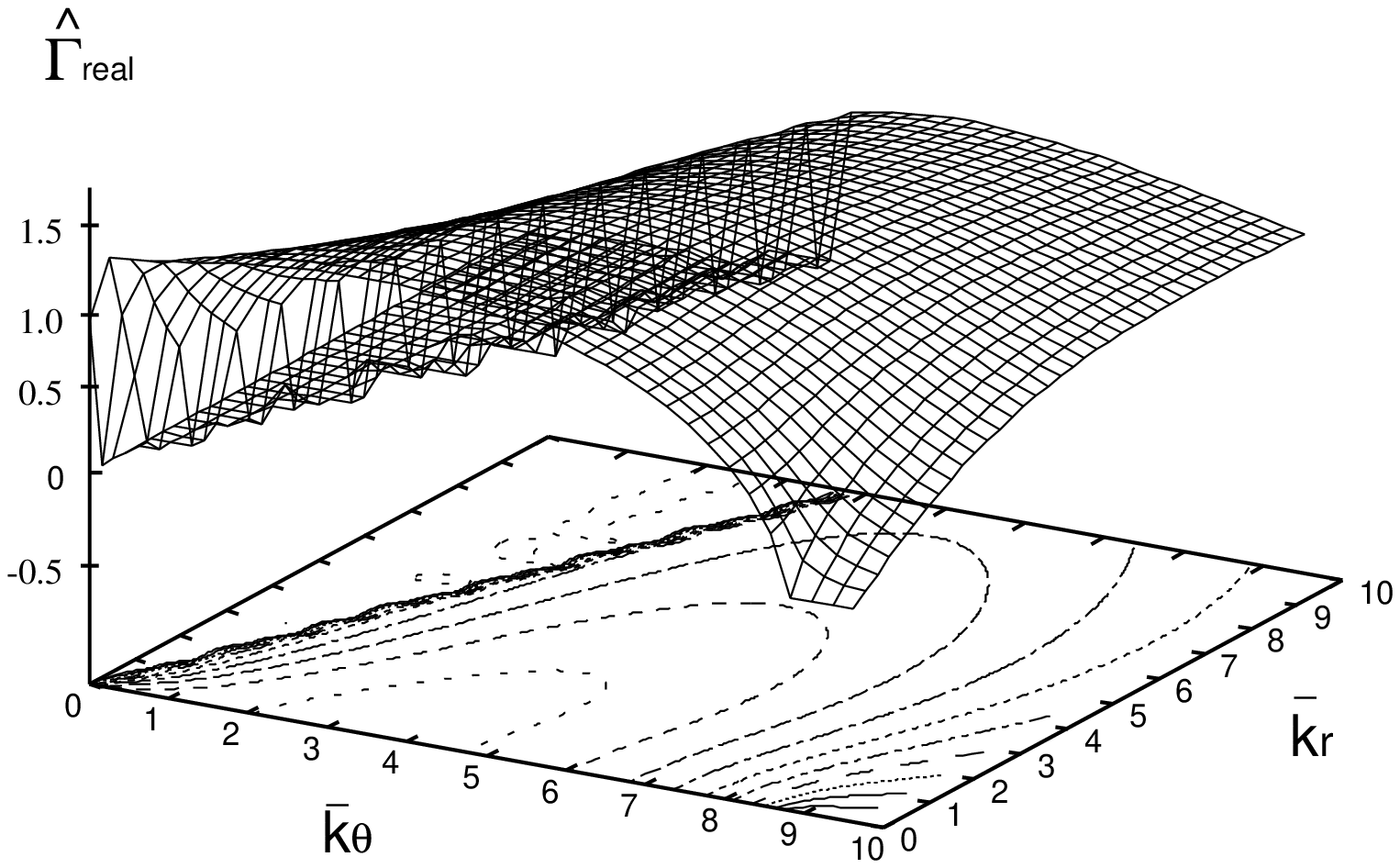}{f12.eps}
\caption{The left panel shows the three-dimensional plot of the unstable
 branch of the dispersion relation with $v_=0.5c$. The right panel shows
 the two-dimensional plot of the imaginary component of the growth  rate
 ${\bar\Gamma}_{Im} (k_{\hat\theta}, k_{\hat r})$. Others are same in
 Fig. 6.  
\label{fig7}}
\end{figure}

\begin{figure}
\figurenum{8}
\plottwo{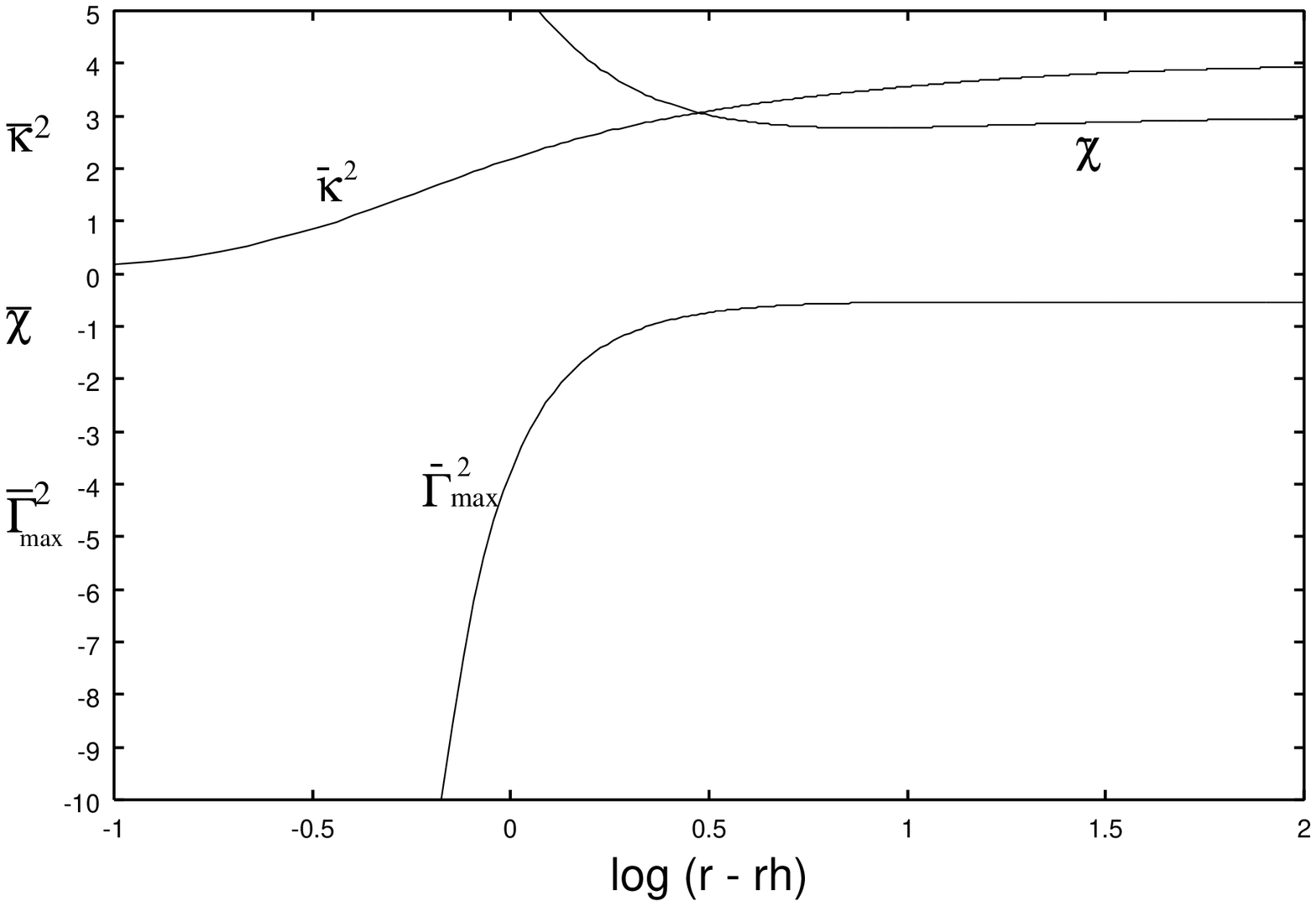}{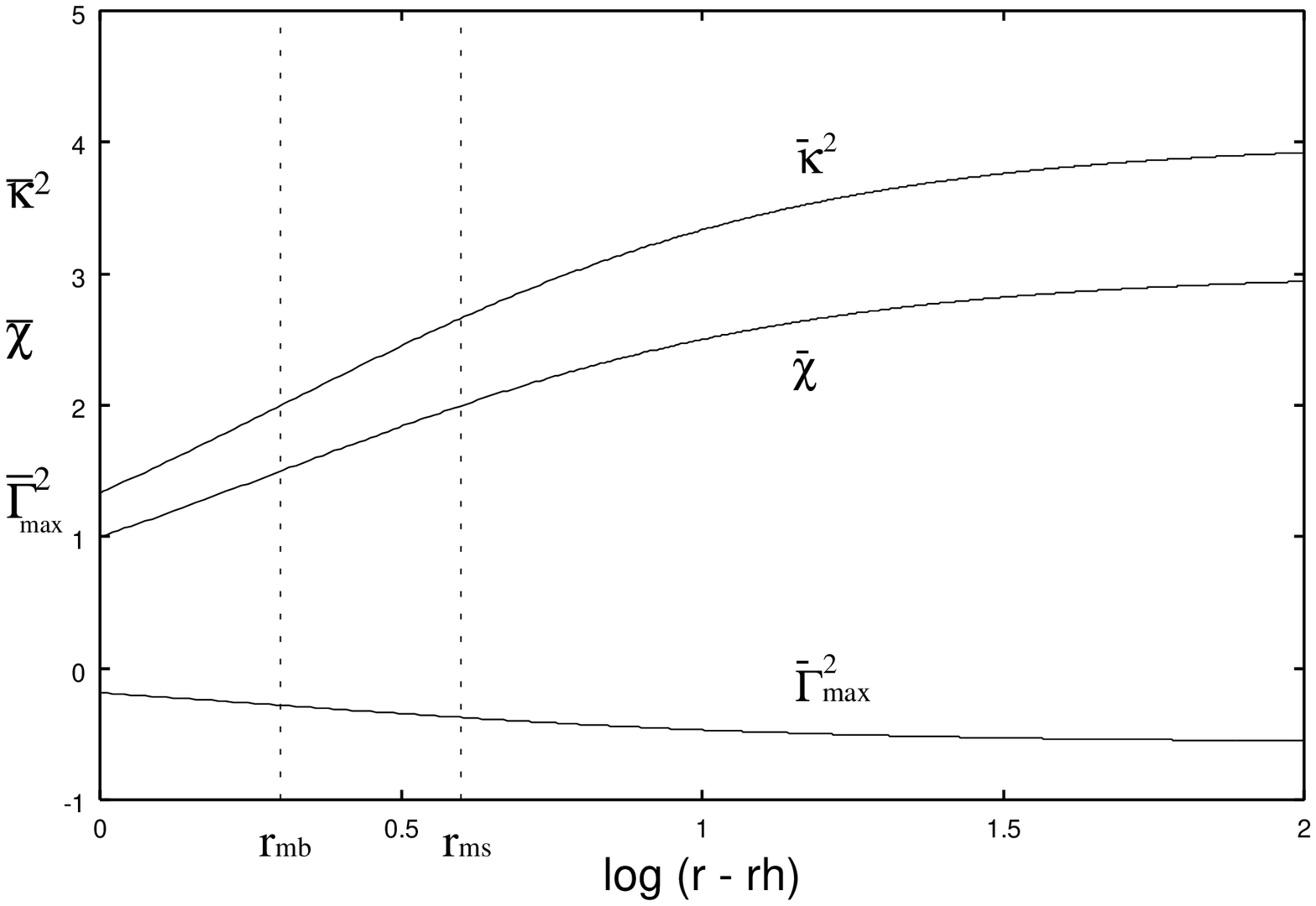}
\caption{The distributions of the maximum growth rate ${\bar\Gamma}_{max}(r)$ and
 the related terms $\bar\chi^2(r), {\bar\kappa}^2(r)$ for the dynamical shear instability at the distance from the event horizon:  The dispersion relation is driven in LNRF. The left figure is shown for $a=M$ and the right one is for $a=0$. 
\label{fig8}}
\end{figure}

\begin{figure}
\figurenum{9}
\plotone{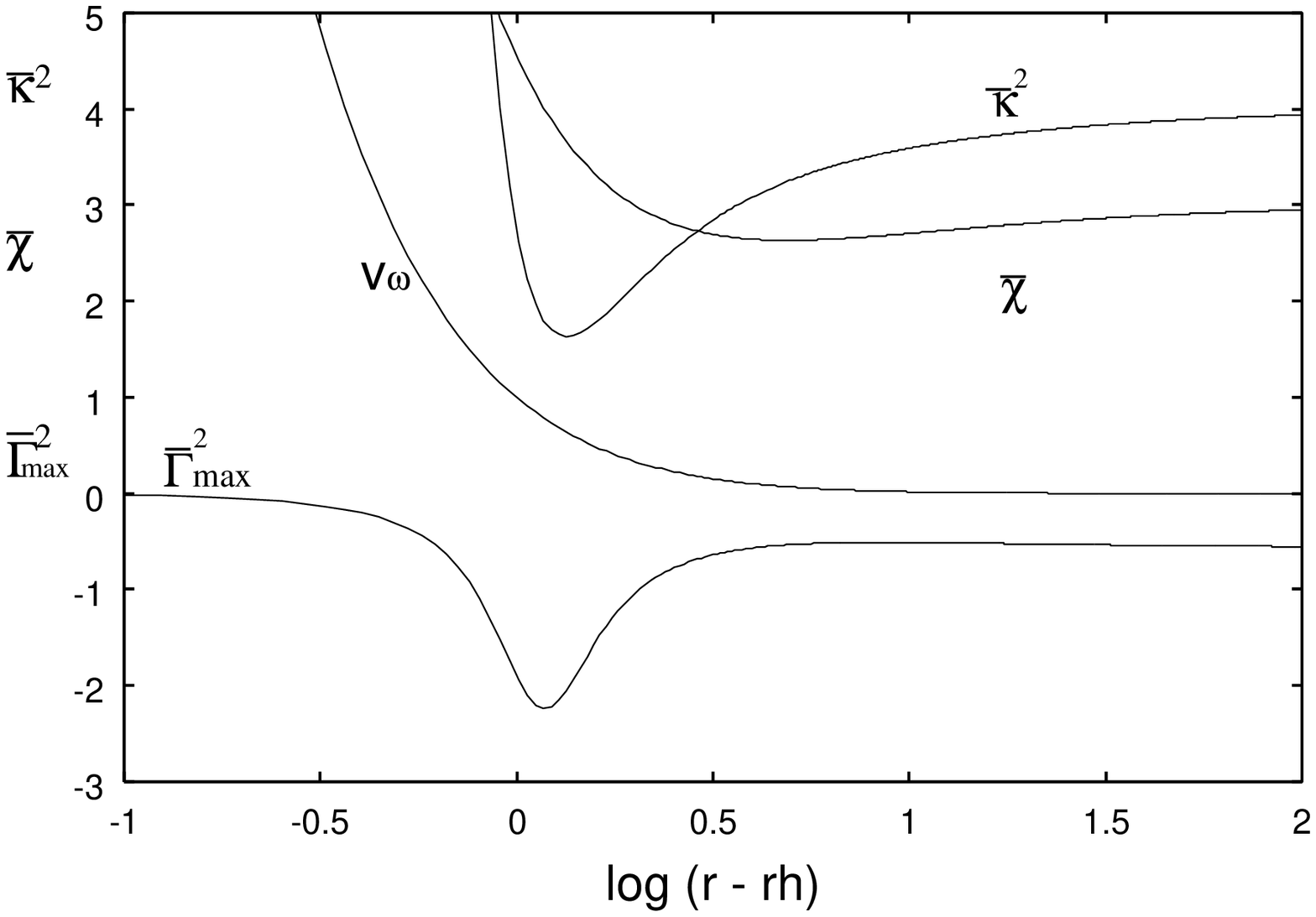}
\caption{The distributions of the maximum growth rate ${\bar\Gamma}_{max}(r)$ and
 the related functions $\bar\chi^2(r), {\bar\kappa}^2(r), v_\omega(r)$ in the dynamical shear instability expressed in the Boyer-Lindquist coordinate frame.
The metric of spacetime is the extreme Kerr.
\label{fig9}}
\end{figure}

\end{document}